\documentclass[sn-nature]{sn-jnl}

\usepackage{graphicx}%
\usepackage{float}
\usepackage{array}%
\usepackage{multirow}%
\usepackage{amsmath,amssymb,amsfonts}%
\usepackage{amsthm}%
\usepackage{mathrsfs}%
\usepackage[title]{appendix}%
\usepackage{xcolor}%
\usepackage{textcomp}%
\usepackage{manyfoot}%
\usepackage{booktabs}%
\usepackage{algorithm}%
\usepackage{algorithmicx}%
\usepackage{algpseudocode}%
\usepackage{listings}%

\usepackage{comment}%

\usepackage{booktabs}%
\usepackage{cuted}%
\usepackage{caption}%
\usepackage[table]{xcolor}%

\usepackage[numbers,sort&compress]{natbib}

\usepackage{needspace,etoolbox}
\makeatletter
\pretocmd{\bibitem}{\Needspace{4\baselineskip}}{}{}
\makeatother

\usepackage[normalem]{ulem}
\AtBeginEnvironment{thebibliography}{\normalem}

\begin{document}

\title[Article Title]{Scalable Reactive Atomistic Dynamics with GAIA}


\author[1]{\fnm{Suhwan} \sur{Song}}\email{suhwan.song@samsung.com}
\equalcont{These authors contributed equally to this work.}

\author[1]{\fnm{Heejae} \sur{Kim}}\email{heejaeee.kim@samsung.com}
\equalcont{These authors contributed equally to this work.}

\author[1]{\fnm{Jaehee} \sur{Jang}}\email{jaehee5.jang@samsung.com}
\author[1]{\fnm{Hyuntae} \sur{Cho}}\email{robert.cho@samsung.com}
\author[1]{\fnm{Gunhee} \sur{Kim}}\email{ghij.kim@samsung.com}
 
\author*[1]{\fnm{Geonu} \sur{Kim}}\email{geonu.kim@samsung.com}

\affil[1]{\orgdiv{Materials AI Lab}, \orgname{AI Center, Samsung Electronics}, \orgaddress{\street{130 Samsung-ro}, \city{Suwon}, \postcode{16678}, \country{Republic of Korea}}}



\abstract{
Groundbreaking advances in materials and chemical research 
have been driven by the development of atomistic simulations~\cite{MRRT53,R64,BBOS83,VDLG01,SMLP10,JEPG21}.
However, the broader applicability of atomistic simulations remains limited, 
as they inherently depend on energy models 
that are either approximate or computationally prohibitive for large-scale simulations.
Machine learning interatomic potentials (MLIPs) have recently emerged 
as a promising class of energy models, but their deployment also remains challenging 
due to the scarcity of systematic protocols for generating training data
spanning diverse structural regimes.
Here we introduce GAIA, an end-to-end automated framework 
that streamlines dataset construction 
for the development of general-purpose reactive MLIPs.
GAIA combines a metadynamics-based exploration scheme
with closed-loop data expansion
for the efficient sampling of a broad spectrum of atomic arrangements,
thereby addressing the reliance on heuristics in conventional dataset generation.
Using GAIA, we constructed Titan25, a benchmark-scale dataset, 
and trained an MLIP that closely matches 
both static and dynamic density functional theory results.
The resulting model reproduces key experimental observations 
across distinct modes of reactivity, 
including detonation, coalescence, and catalytic processes.
GAIA thus helps bridge the gap between simulation and experiment, 
paving the way toward scalable and general MLIPs 
capable of describing a wide range of materials and chemical processes.
}

\maketitle

\section{Introduction}

Atomistic simulations are advancing 
the frontiers of materials and chemical sciences.
This progress has been driven by simulation techniques 
such as molecular dynamics (MD) and Monte Carlo methods, 
which have been pivotal in elucidating phenomena 
that remain inaccessible to direct experimental observation~\cite{CP85,FW91,IMT05,MH09,HD18}.
However, the fidelity of these simulations 
depends on the underlying energy models, 
which involve inevitable trade-offs 
between predictive accuracy and computational efficiency. 
Classical force fields (FFs) are widely used energy models that approximate complex atomistic interactions 
with simple analytical functions, 
thereby providing the computational efficiency required for large-scale simulations~\cite{PS17,SMDS19}.
Nevertheless, the reliance on heuristics and system-specific parameter fitting 
limits their general applicability, 
particularly for systems involving transition metals and complex chemistries~\cite{HSMM18}.
This limitation is compounded by the fact that
relatively few FFs can describe bond formation and breaking, 
an essential capability for modeling complex chemical processes. 
As a result, their use is often restricted to 
narrowly defined systems and applications~\cite{STH00,YSP07,SHIK16}.
Density functional theory (DFT) can address these challenges
by providing an accurate description of interatomic interactions, 
but its applicability is limited to simplified systems 
rather than experimentally relevant ones,
since the computational cost increases significantly with system size.

Recently, machine learning interatomic potentials (MLIPs) have emerged 
as a promising approach for large-scale simulations~\cite{JMAM25}.
By learning from high-quality reference data,
MLIPs can approach ab initio-level accuracy
while exhibiting linear computational scaling.
As a result, 
they have attracted considerable attention 
as a route to narrowing the gap between computation and experiment
in atomistic simulations~\cite{WNWB22,LBFQ23}.
However, in the absence of sufficient and varied training data,
the performance of MLIPs deteriorates, 
limiting their reliability.
This limitation, 
coupled with the scarcity of general-purpose MLIPs
capable of handling diverse chemical environments with a single model,
highlights the need for comprehensive protocols for generating 
a wide spectrum of representative atomic configurations.

In this paper, we introduce GAIA, 
an automated framework for constructing reactive MLIP datasets.
GAIA is designed to address the aforementioned challenges---namely, 
the lack of systematic strategies for sampling structural space across multiple chemical domains
and the difficulty of ensuring transferability among them---thereby 
enabling a broad range of simulations with a single MLIP model.
To this end, 
GAIA implements an end-to-end workflow 
that generates a rich ensemble of atomic configurations for MLIP training,
starting from simple, user-specified seed structures.

Using GAIA,
we construct Titan25, a dataset
comprising 1.8 million atomic configurations across eleven elements.
We show that SNet-T25, an MLIP trained on Titan25,
achieves close agreement with DFT and experimental results
for metal–organic adsorption distances, 
coalescence of carbon nanotubes, 
detonation product distributions, 
and the water–gas shift reaction mechanism (Fig.~\ref{fig:fig1}a,b).
We also find that SNet-T25 outperforms recent state-of-the-art models 
on a suite of out-of-distribution benchmarks.
Taken together, these results demonstrate that 
GAIA can serve as a general framework 
for developing transferable MLIPs across diverse, chemically reactive systems.

\section{Toward generalizable MLIPs} 

\begin{figure}[ht]
    \centering
    \includegraphics[width=1.0\textwidth]{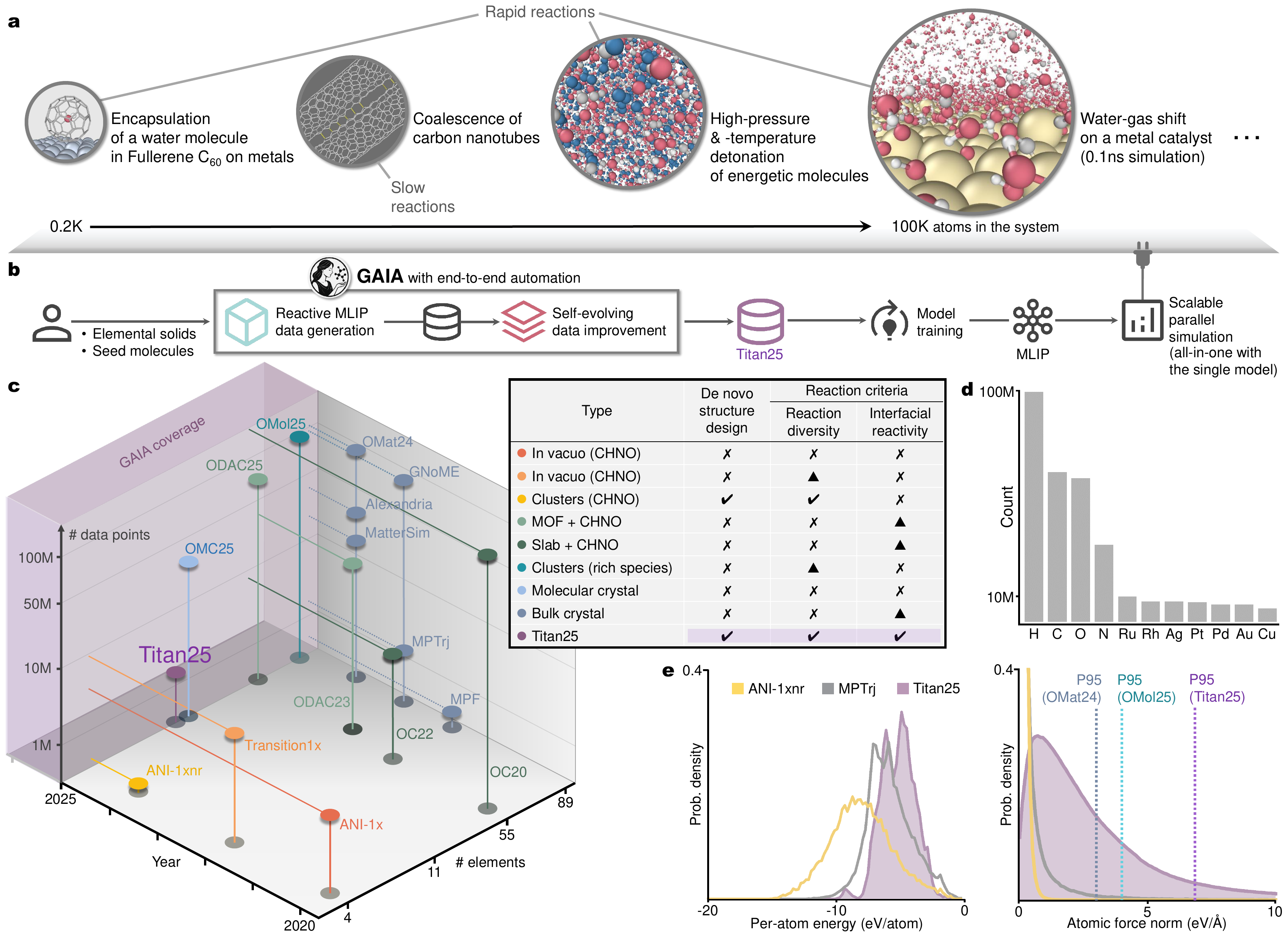}
    \caption{\small
        \textbf{Large-scale atomistic simulations with the Titan25-trained MLIP.}
        \textbf{a},~MLIPs have so far shown limited transferability across diverse reactions,  
making it difficult to apply them broadly as general models.
Here, we show representative simulations enabled by a single Titan25-trained model, spanning four scales from left to right: 
molecular adsorption, nanoscale coalescence, non-equilibrium assembly, and catalytic reactions.
Together, these examples demonstrate the broad applicability of the MLIP.
        \textbf{b},~Schematic overview of the GAIA framework for generating training datasets and deploying MLIPs. 
From left to right, user-provided input structures are augmented by GAIA into diverse datasets 
(for example, Titan25) comprising a wide range of atomic arrangements. 
These datasets can then be used to train MLIPs, which are subsequently applied in the simulations.
        \textbf{c},~Comparison of dataset scales, with axes 
showing the release year, the number of elements, and the number of data points. 
GAIA can generate datasets with scalable sizes and elemental diversity. 
As a representative case, Titan25 comprises eleven elements and 1.8 million data points.
An inset table qualitatively summarizes whether each compared dataset includes de novo configurations
and is capable of describing diverse reactions and interfacial processes.
Further details of this summary are provided in Supplementary Note~\ref{supp:related_work}. 
        \textbf{d},~Element-wise atom counts in Titan25.
        \textbf{e},~Probability density distributions of per-atom energies (left panel) and norms of atomic forces (right panel).
P95 denotes the 95$^\text{th}$ percentile. See Supplementary Fig.~\ref{sfig:titan25} for additional details on Titan25.
}
    \label{fig:fig1}
\end{figure}

The generalization of MLIPs critically hinges on training data that encompass structural diversity~\cite{KNLS24}. 
In principle, a hypothetical dataset that fully covers the entire chemical space would resolve this limitation, 
yet constructing such a dataset is infeasible in practice, 
leaving generalization as a persistent challenge. 
State-of-the-art approaches are often constrained---either 
by limited structural diversity due to incremental refinements of existing databases or 
by substantial overhead that extends well beyond data generation itself (Supplementary Note~\ref{supp:related_work}). 
To address these limitations, we designed GAIA to generate datasets 
that broadly represent the accessible chemical space spanned by user-chosen seed components, 
while reducing redundancy and promoting transferability across distinct chemical environments. 
The framework thus allows datasets to scale systematically across the targeted chemical space 
as the number of seed components increases. 
Titan25, a representative dataset generated by GAIA,
comprises a moderate, million-scale number of data points,
yet it includes de novo configurations suitable for describing a wide range of chemical reactions (Fig.~\ref{fig:fig1}c)
and exhibits a well-balanced distribution of elements (Fig.~\ref{fig:fig1}d).
The distribution of per-atom energies in Titan25 is shifted toward higher energy values 
relative to ANI-1xnr~\cite{ZMJK24} and MPTrj~\cite{DZJR23}, 
and the 95$^\text{th}$ percentile of atomic force norms reaches 7.1\,eV/\,\AA, 
exceeding those of not only ANI-1xnr and MPTrj but also OMat24~\cite{BSFW24} and OMol25~\cite{LSST25} 
(Fig.~\ref{fig:fig1}e and Supplementary Fig.~\ref{sfig:titan25}c,d). 
These statistics indicate that Titan25 spans a broader spectrum of energetic and force regimes, 
encompassing various near- and non-equilibrium structures. 
Such coverage provides informative training signals for MLIPs, 
particularly in reactive regimes, and positions Titan25 as a prototype dataset 
that can be readily extended by GAIA to broader regions of chemical space.

\section{GAIA framework} 

The GAIA framework aims to provide comprehensive training data that enable a single MLIP model 
to generalize across crystalline, amorphous, molecular, and interfacial systems. 
It consists of two main modules: 
the data-generator (DG), which produces a broad spectrum of atomic arrangements 
by introducing multiple physics-inspired structure builders, and 
the data-improver (DI), which augments them through coverage-enhancing sampling of structural space. 
The DG module generates DFT-labeled snapshots from simple user-provided seed components---such as 
H$_2$O, CO$_2$, Pt(111)---using six builders: Checkerboard, Bulk, Slab, Adatom, Admol, and Nanoreactor$^+$ (Fig.~\ref{fig:fig2}a). 
The first five builders generate structures by varying the input seed components according to predefined protocols. 
In contrast, Nanoreactor$^+$ systematically uses metadynamics to explore chemical transformations, 
producing non-equilibrium data points that are crucial for describing reactions involving metals and nonmetals. 
Together, these builders generate periodic bulk, surface, mixed-composition, and reactive configurations, 
thereby expanding the diversity of the training data 
(Methods~\ref{method:dg} and Supplementary Figs.~\ref{sfig:dg}~and~\ref{sfig:nano}). 
The clustered embedding observed in Fig.~\ref{fig:fig2}b highlights the complementary sampling patterns of different builders and 
shows that structures generated by Nanoreactor$^+$ occupy regions of structural space distinct from those of the other builders. 
The DI module generates additional training data, 
targeting regions that are underrepresented in the DG-generated set or 
in which the trained model exhibits high prediction errors (Fig.~\ref{fig:fig2}c). 
It serves as a more scalable alternative to existing approaches 
that rely on dense sampling of candidate structures~\cite{BCD19,LMEF25} or on the guidance of iteratively trained models~\cite{KBLL23,OSKO23,ZMJK24}. 
The DI module identifies the improvable regions by performing data categorization in a two-dimensional space 
spanned by the total energy and the atom–atom distance diversity $d_{\Lambda}$, which quantifies diversity across atomic configurations (Methods~\ref{method:di}). 
Fig.~\ref{fig:fig2}d visualizes this space, comparing the distributions of structures generated by the DG and DI modules. 
The DI module extends sampling into regions scarcely populated by DG, as seen in the marginal distributions, 
thereby enhancing the coverage of configuration space. 
Both DG and DI are automated and require only minimal domain knowledge, 
offering greater scalability than conventional manual curation or other computationally costly workflows 
that can yield MLIP datasets with relatively limited scope.
Further methodological comparisons with prior work are provided in Supplementary Note~\ref{supp:related_work}.

\section{GAIA benchmark suite}

To quantify the predictive performance of MLIPs for fundamental physical properties,
we constructed GAIA-Bench, a benchmark suite providing DFT reference energies and forces 
for four tasks---intermolecular interactions (mol2mol), bulk energy–volume relations (bulk), surface facet stability (slab), and molecule–surface adsorption energetics (mol2surf).
In constructing GAIA-Bench, the benchmark structures were rigorously selected to avoid overlap with the training datasets compared here,
enabling an assessment of generalization that is not unduly biased toward any particular one.
For the force evaluations, perturbed configurations of the backbone structures were additionally considered
at multiple levels ($\epsilon$) to probe a wider range of force variations.
Further details of GAIA-Bench are provided in Methods~\ref{method:gb}.
Using GAIA-Bench, we evaluated four MLIPs:
two based on GAIA-constructed datasets,
Titan25(G) (from DG only) and Titan25(G+I) (from both DG and DI),
and two counterparts trained on the existing public datasets, ANI-1xnr and MPTrj.
For a fair comparison, all four models shared the same architecture and training protocol (Methods~\ref{method:mt}).
Under these conditions, the model trained on Titan25(G+I) consistently achieved the lowest errors across all GAIA-Bench tasks.
Its energy errors are approximately halved for mol2mol and mol2surf relative to the public-data models,
while its force errors are reduced by about one-third on average across the four tasks.
Relative to Titan25(G), Titan25(G+I) also shows notable performance gains in most tasks and metrics (Fig.~\ref{fig:fig3}a).
The force results in Fig.~\ref{fig:fig3}b show that
the shaded error regions grow markedly for the public-data models as $\epsilon$ increases, 
whereas they remain relatively narrow for the Titan25-based models.
Moreover, Titan25(G+I) yields the smallest errors for both the magnitude and angular components of the forces (Fig.~\ref{fig:fig3}c).
These results indicate that training data generated by DG already deliver strong generalization across diverse configurations,
and that integrating DG with DI further amplifies this,
supporting progress toward broadly transferable MLIPs.
Detailed numerical values and extended evaluations are reported in 
Supplementary Figs.~\ref{sfig:bench}~and~\ref{sfig:parity_bench}, and Supplementary Tables~\ref{stab:bench_e2f}~and~\ref{stab:public_test}.

\section{Reproducing prototypical surface reactions: Pt--CO$_x$--H$_2$O}

Alongside the benchmark evaluation of static energies and forces,
we further assess how well MLIPs reproduce interfacial dynamics
through MD simulations of CO and CO$_2$ adsorption on Pt(111) in aqueous solution~\cite{LHI18, ZKCW19, SYLZ20, GS22}. 
In these evaluations, the SevenNet model~\cite{PKHH24} trained on the Titan25(G+I) dataset, termed SNet-T25,
is compared with UMA~\cite{WDFG25}, a massive model jointly trained on five large-scale MLIP datasets (OC20, ODAC, OMat, OMol, and OMC).
The details of the simulation setups and the models are described in Methods~\ref{method:ptcox}.
As shown in Fig.~\ref{fig:fig3}d, SNet-T25 and the three UMA variants (UMA-OC20, -ODAC, and -OMat)
consistently reproduced the Pt--CO adsorption distance ($\sim$1.4\,\AA), in close agreement with DFT ($\sim$1.45\,\AA).
In contrast, UMA-OMol and -OMC, which are specialized for molecular data, 
exhibited instabilities in the Pt surface structure.
For the CO$_2$ adsorption case, only SNet-T25 maintained the molecules close to the Pt surface ($\sim$3.4\,\AA), 
whereas the UMA variants showed large fluctuations, indicative of repulsive interactions (Fig.~\ref{fig:fig3}e).
We also performed DFT calculations on reduced Pt--CO$_2$ systems (Methods~\ref{method:ptcox}), 
which provided distance values consistent with those predicted by SNet-T25.
Overall, the results demonstrate that SNet-T25 leads to accurate representation of both CO and CO$_2$ adsorption, 
outperforming UMA on this aqueous interface. 
This highlights the ability of Titan25-trained MLIPs 
to reproduce realistic interfacial dynamics with near-ab initio fidelity, 
capturing the intricate behavior of molecules at solid--liquid boundaries in chemically complex environments, 
even though SNet-T25 is markedly smaller in model size
and is trained on considerably less data than UMA.

\section{Validation across experimental phenomena}

Beyond achieving DFT-level accuracy, 
a central practical value of MLIPs is their capacity to reproduce experimental observables. 
However, achieving such agreement for systems not explicitly represented in their training domain
remains a challenge.
In the following subsections, we examine the generality of a single MLIP, SNet-T25,
across a range of experimentally relevant chemical processes.
Using this potential, we simulate detonation, coalescence, adsorption, and catalytic mechanisms, 
spanning from organic systems to metal--molecule interfacial processes.
See Methods~\ref{method:MD} for detailed simulation conditions.

\subsection{Detonation of energetic molecules}

Detonation is a shock-driven combustion process at extreme pressures and temperatures, 
where accurate prediction of product distributions is essential for assessing energy release and validating models~\cite{GZAG16,HYSI23}.
We performed Hugoniostat simulations at 40~GPa for TNT, TATB, RDX, and nitromethane using SNet-T25 (Fig.~\ref{fig:fig4}a), 
and compared the normalized product ratios with experimental data~\cite{O82} and with the domain-specific MLIP, Gen3.9zbl~\cite{HYSI23}.
As shown in Fig.~\ref{fig:fig4}b, SNet-T25 predicts the formation of H$_2$O, CO$_2$, N$_2$, and NH$_3$ 
with normalized ratios comparable to both experiment and Gen3.9zbl for all four materials.
The deviations from experiment are the smallest for nitromethane, remaining below 4\% across all products, 
and the largest for TATB, reaching 18\% for N$_2$ due to the overproduction.
SNet-T25 also reproduces the experimental ordering of product distributions across all systems,
except for TNT, for which the three major products are obtained in nearly identical ratios (within $\sim$2\% deviation).
Supplementary Fig.~\ref{sfig:deto} shows that SNet-T25 yields consistent product ratios across multiple pressures, 
indicating that this Titan25-trained MLIP remains numerically stable and chemically reasonable under these high-pressure detonation conditions, 
despite not using additional short-range correction terms such as the Ziegler–Biersack–Littmark (ZBL) potential~\cite{ZB85,HYSI23}.

\subsection{Coalescence of carbon nanotubes}

The coalescence of carbon nanotubes (CNTs) provides a route to tune their diameter and morphology, 
but simulating this process has been difficult 
because ab initio molecular dynamics cannot reach the relevant spatiotemporal scales, 
while FFs struggle to capture bond rearrangements~\cite{TTBC00,TNHC25}.
Using SNet-T25, we simulated the coalescence of pairs of armchair CNTs 
with chiral indices (6,6), (9,9), (12,12), (18,18), and (24,24) (Fig.~\ref{fig:fig4}c--e). 
For the (6,6), (9,9), and (12,12) pairs, 
the resulting coalesced structures exhibited pore diameters consistent with those of the corresponding doubled tubes, 
(12,12), (18,18), and (24,24), respectively, 
in agreement with experimental observations of (n,n) $\rightarrow$ (2n,2n) coalescence.
By contrast, a pair of (24,24) tubes collapsed into ribbon-like structures rather than forming a (48,48) cylinder, 
also consistent with the experimental trend that large-diameter CNTs ($\approx$\,4.2\,–\,6.9\,nm) 
lose radial stability and flatten upon contact~\cite{ESWY04,LKHM22}.
These results indicate that SNet-T25, trained on Titan25, 
can reliably capture bond rearrangements in extended carbon systems,
enabling predictive simulations of size- and morphology-dependent CNT behavior.

\subsection{$\pi$-conjugated molecules on metal surfaces}

The equilibrium adsorption distance of $\pi$-conjugated molecules on metals is a key descriptor of hybrid interface stability, 
yet its experimental determination remains restricted to highly ordered overlayers 
and requires complex analysis~\cite{WHHB91,Z93,WB94,W05,LPHW09,BFTG13,SSDP14,LMWB15}.
Fig.~\ref{fig:fig5}a,b show adsorption distances of benzene~(BZ), diindenoperylene~(DIP), and fullerene~(C$_{60}$) on Ag(111),
as obtained from experiment~\cite{MRCL16}, DFT~\cite{MRCL16}, and SNet-T25.
Considering all examined metal substrates (Ag, Au, Cu, and Pt),
SNet-T25 reproduces the parallel adsorption geometries 
with a mean absolute relative deviation of $\approx 6$\% from experiment across the three $\pi$-conjugated adsorbates  
(Supplementary Fig.~\ref{sfig:pi-conj-relaxation}).
For DIP, the mean deviation is slightly larger ($\approx 7$\%), 
yet the model correctly captures the relative order of experimental adsorption distances, Cu $<$ Ag $<$ Au.
A more complex case is water encapsulated inside C$_{60}$ on Ag(111)~\cite{JSJG21}, 
where SNet-T25 reproduces the experimental Ag--O distance (5.60\,\AA\, vs. 5.57\,\AA) 
in the time-averaged MD trajectories (Fig.~\ref{fig:fig5}c).
Together, these results suggest that SNet-T25 can capture the delicate balance of dispersion, orbital hybridization, and intramolecular reorganization 
that governs $\pi$-conjugated molecules on metal surfaces, 
providing a physically grounded framework for interpreting and predicting interfacial stability. 

\subsection{Water-gas shift reaction on Au(100)}

The water–gas shift reaction (CO + H$_2$O $\rightarrow$ CO$_2$ + H$_2$) is a key route to hydrogen production, 
yet its catalytic mechanism remains contested, 
as the activity is highly complex and sensitive to subtle variations in reaction conditions~\cite{RW09,EKK20}.
Using SNet-T25, we performed large-scale MD simulations of CO and H$_2$O on hydroxylated Au(100) (Fig.~\ref{fig:fig5}d). 
The trajectories capture the essential features of two pathways.
In the carboxyl mechanism, CO adsorption is followed by the formation of the OCOH* intermediate and subsequent CO$_2$ release mediated by adjacent H$_2$O and OH.
In the redox mechanism, H-abstraction yields O*, which couples with CO* to form CO$_2$. 
These results demonstrate that SNet-T25 can reproduce experimentally established reaction channels
and resolve their dynamics under realistic catalytic conditions,
establishing GAIA-constructed datasets as a robust foundation for predictive modeling of heterogeneous catalysis.

\section{Conclusion}

We have demonstrated that
GAIA can systematically construct diverse, high-quality datasets,
enabling MLIPs to describe not only static energies and forces 
but also dynamical chemical processes, including reactions in molecular, solid, and interfacial systems.
This is exemplified by Titan25, a benchmark-scale dataset comprising 1.8 million structures,
and by SNet-T25, a Titan25-trained model that exhibits broad expressivity across various reactive regimes.
A notable advantage of GAIA is that
it enables end-to-end automated generation of MLIP training datasets,
thereby substantially alleviating reliance on expert heuristics 
traditionally required to manually design and curate training data.
This data-centric strategy leads to robust performance on out-of-distribution structures,
as demonstrated on the GAIA-Bench suite,
where the models trained on the GAIA-generated datasets
retained their accuracy even for perturbed and previously unseen configurations.
In particular, the SNet-T25 model reproduced experimental observables
ranging from detonation products to catalytic pathways,
showing its transferability across these distinct chemical domains.
These results suggest that
GAIA may facilitate the evolution of MLIPs into predictive engines 
for exploring materials and chemical space under realistic conditions,
including regimes that are challenging for conventional simulations.

The configuration space of atomic systems is in principle infinite, 
raising the question of whether a truly single-model universal MLIP is achievable
or remains an aspirational ideal. 
In practice, efforts to develop such a model have been hindered 
by a lack of generally applicable, practically scalable approaches for identifying
which regions of chemical space are adequately sampled and which remain underrepresented, 
rendering the construction of exhaustive datasets effectively prohibitive.
Nevertheless, by systematically and automatically generating datasets 
initiated from simple seed components and
progressively expanded into neighbouring regions of chemical space, 
GAIA provides a practical foundation for the development of 
increasingly general, potentially universal MLIPs.


\clearpage

\begin{figure}[ht]
    \centering
    \includegraphics[width=1.0\textwidth]{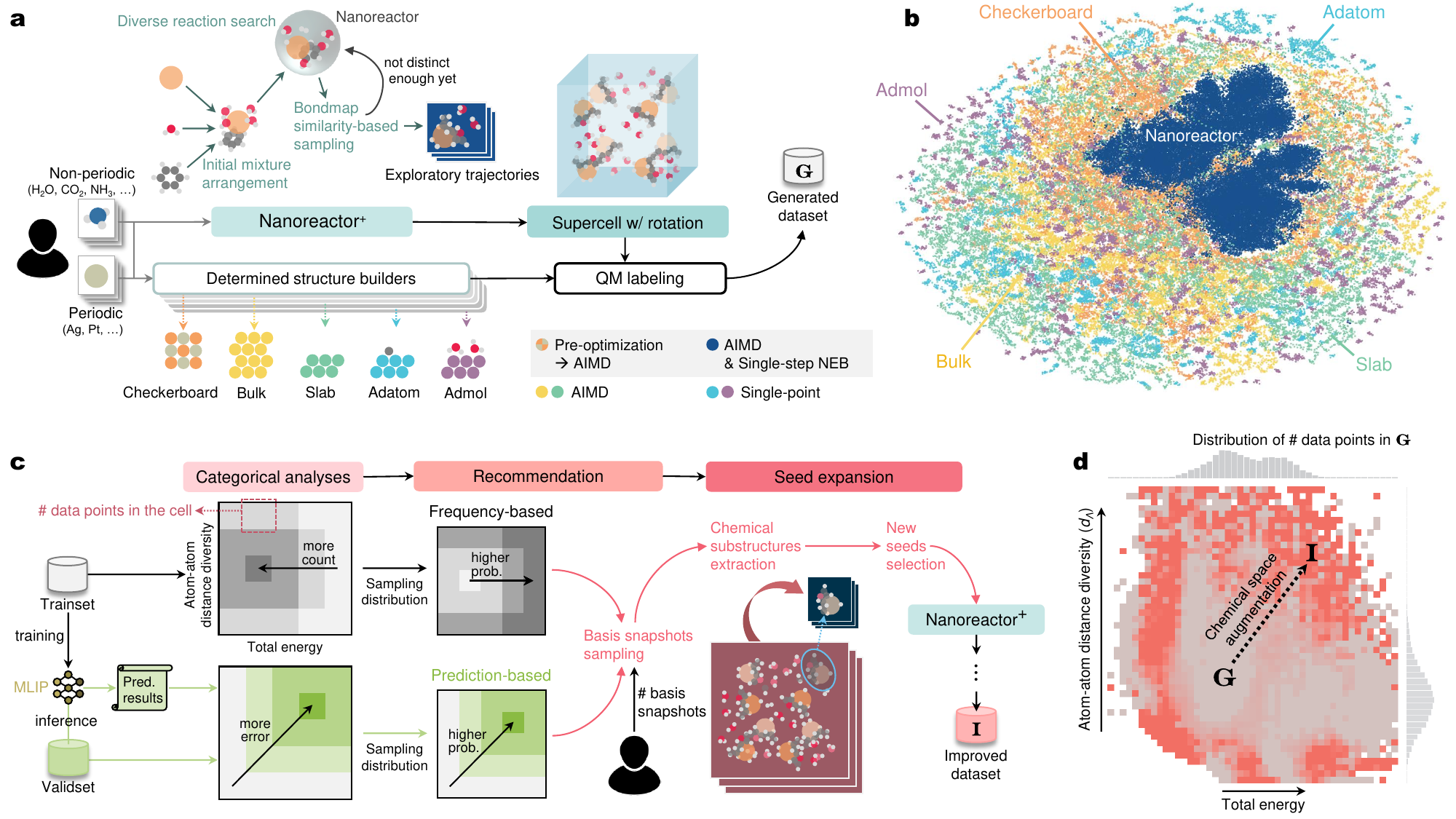}
    \caption{\small
        \textbf{GAIA data-generator and -improver produce diverse datasets.} 
        \textbf{a},~Illustration of the DG module. 
Simple periodic and non-periodic input components are expanded into diverse unlabeled configurations by builders 
such as Checkerboard, Bulk, Slab, Adatom, Admol, and Nanoreactor$^+$. 
The first five builders generate structures by retaining key input features or combining structures, 
while Nanoreactor$^+$ explores reactive configurations by inducing bond formation and cleavage via enhanced dynamics.
The generated structures are periodically replicated to form supercells with random rotations,
which are then labeled at the DFT level.
        \textbf{b},~Feature-space embedding of structures generated by the DG module (dataset G),
obtained from a Titan25-trained MLIP.
Each point corresponds to a generated structure, and the color denotes the builder from which the structure originated.
        \textbf{c},~Illustration of the DI module.
For the exploratory data points, basis snapshots are extracted through categorical analyses that 
respectively rely on data distribution statistics from the training dataset and error-based measures derived from model predictions.
New seed structures selected from these snapshots are fed into the Nanoreactor$^+$ builder, 
where they are expanded into diverse configurations to form dataset I.
        \textbf{d},~Comparison of structures generated by DG and DI 
using atom--atom distance diversity ($d_\Lambda$) and total energy. 
Heatmap elements show the relative dominance of each dataset by counts, with gray indicating dataset G and red indicating dataset I. 
The distribution of dataset G is further illustrated as a histogram on the side.
Red regions correspond to dataset I populating sparse areas of dataset G, 
demonstrating its contribution of additional structures beyond those obtained from DG alone.
    }
    \label{fig:fig2}
\end{figure}

\begin{figure}[ht]
    \centering
    \includegraphics[width=1.0\textwidth]{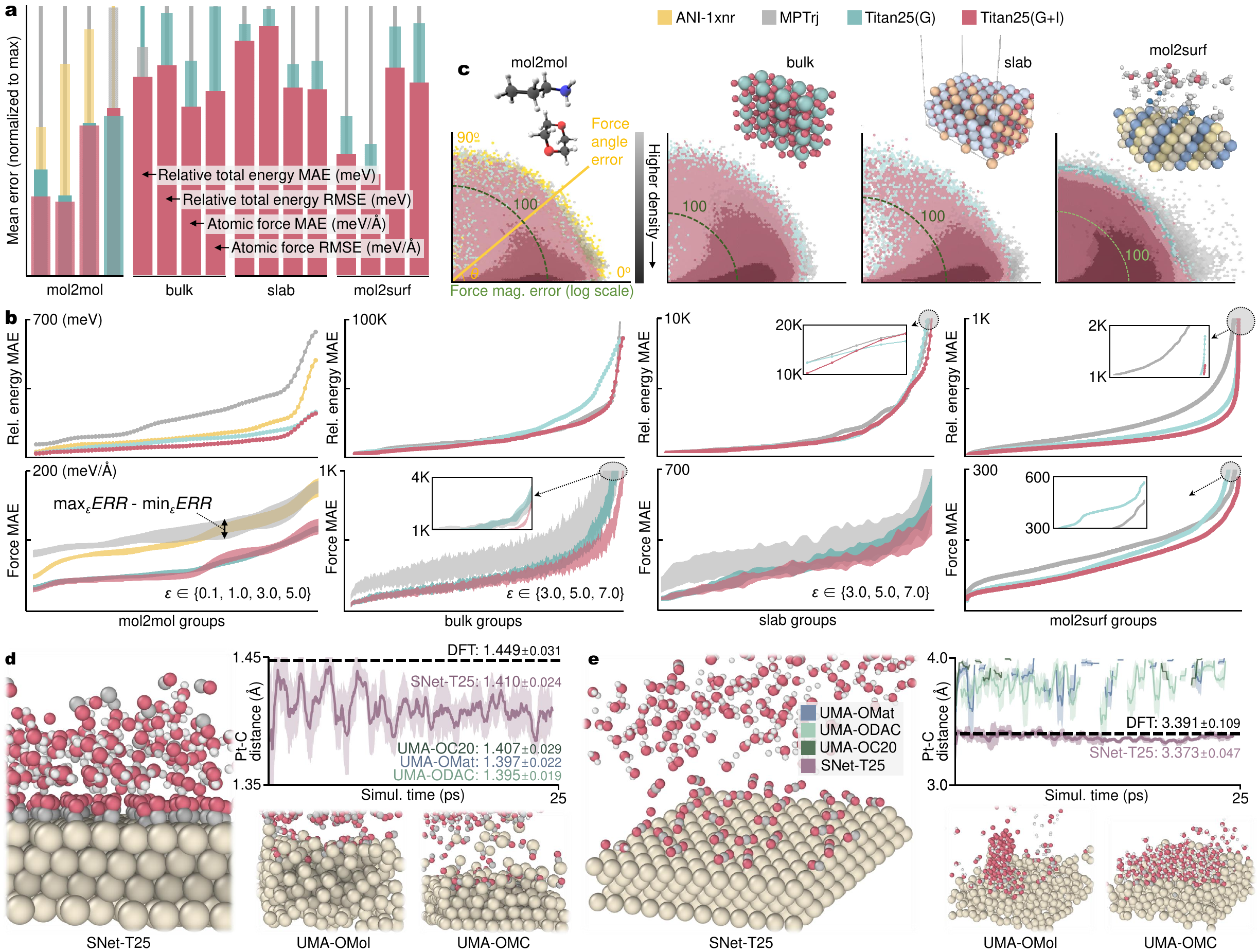}
    \caption{\small
        \textbf{The Titan25-trained MLIP achieves reliable accuracy in both static and dynamic benchmarks.} 
        \textbf{a},~Plot of normalized mean errors on GAIA-Bench, the static benchmark suite. 
The normalization scales all bars to a common maximum for relative comparison.
ANI-1xnr is shown only for the mol2mol task, as it does not include metallic elements.
        \textbf{b},~Mean absolute errors (MAEs) of relative energies (top) and forces (bottom) for the four GAIA-Bench tasks.
Each plot is individually sorted in ascending order; 
each point on the horizontal axis corresponds to a group of snapshots derived from the same reference structure.
In the lower panel, shaded regions represent the range between the smallest and largest MAE values across different values of $\epsilon$, 
where $\epsilon$ denotes the structure perturbation strength at which the force norm matches the threshold.
For mol2surf, no additional perturbations were applied; 
instead, only the structures used for energetic calculations were considered.
An inset provides an enlarged view of the plot near its end.
Root-mean-square error (RMSE) results are provided in Supplementary Fig.~\ref{sfig:brmse}.
        \textbf{c},~Angle and magnitude errors in atomic forces. Magnitude errors are shown with a reference level of 100\,meV/\,\AA.
        \textbf{d},~Evolution of the Pt–C vertical distance 
during MD simulations of CO and H$_2$O adsorbed on the Pt(111) surface using the SNet-T25 model. 
Ivory, gray, white, and red spheres represent Pt, C, H, and O atoms, respectively. 
The moving-average curve is presented for SNet-T25;
for DFT, the mean value is shown together with a horizontal dashed line;
for UMA-OMat, -OC20, and -ODAC, only the mean values are provided.
We denote these models as UMA-X, since the forward path of the UMA model is adapted depending on the task input X.
Snapshots from the simulations with UMA-OMol and -OMC are also displayed, 
illustrating that the Pt surface structure was not preserved.
        \textbf{e},~As~in~\textbf{d}, except that CO is replaced by CO$_2$. 
The distance curves are displayed for the four MLIPs, 
whereas only the mean value is shown with a horizontal dashed line for DFT.
The apparent truncation of the UMA-X curves arises from cases where the Pt--C distance exceeds 4.0~\AA.
    }
    \label{fig:fig3}
\end{figure}

\begin{figure}[ht]
    \centering
    \includegraphics[width=1.0\textwidth]{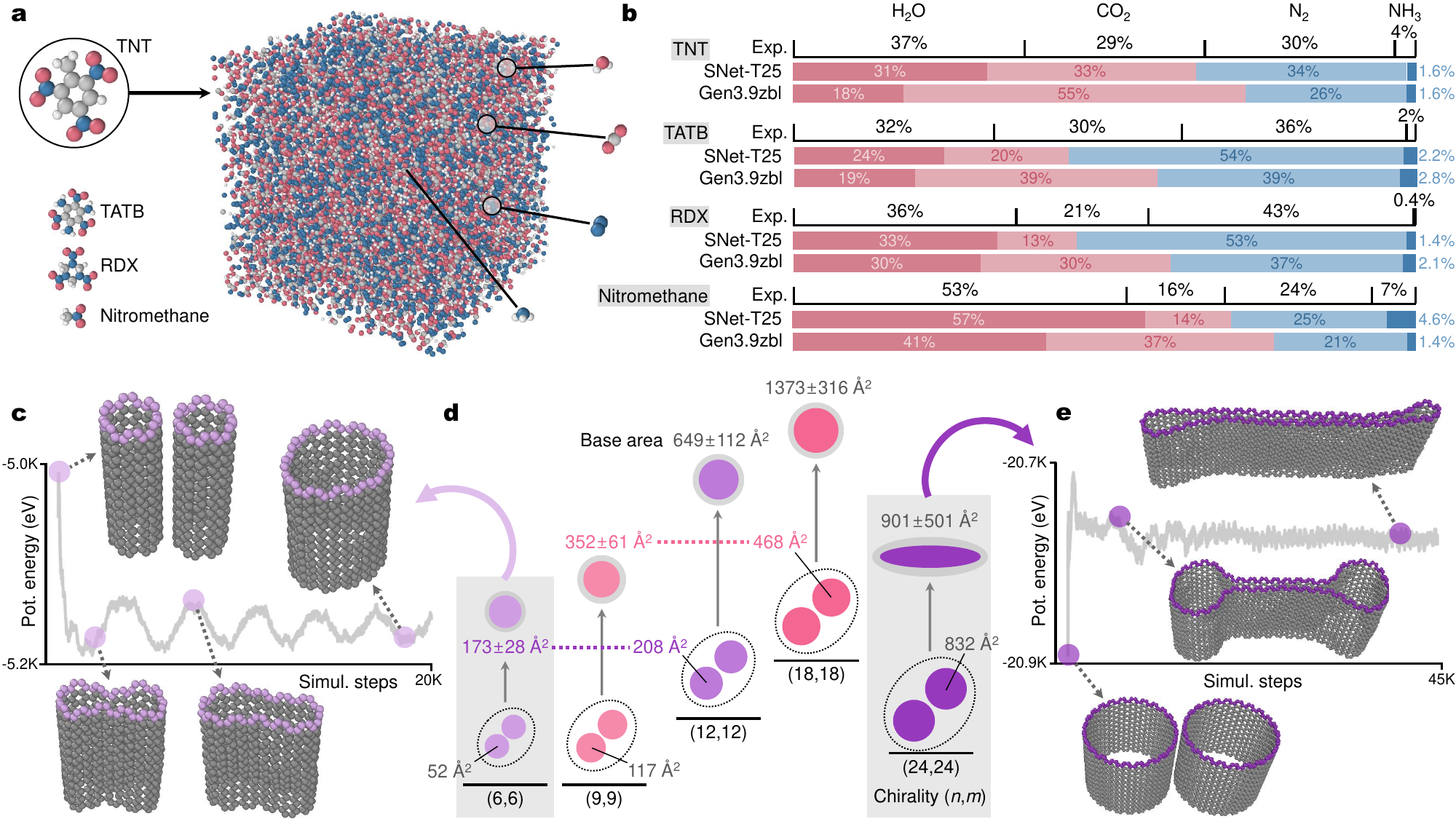}
    \caption{\small
        \textbf{Atomistic simulations under extreme conditions: detonation products and 
        CNT coalescence.} 
        \textbf{a},~Snapshot of 1000 TNT molecules after a 40~GPa Hugoniostat simulation. 
Gray, white, blue, and red spheres represent C, H, N, and O atoms, respectively. 
For reference, single-molecule structures of TATB, RDX, and nitromethane are also shown; 
their post-simulation configurations are not displayed as they appear similar to TNT.
        \textbf{b},~Normalized product distributions after Hugoniostat simulations of each compound. 
Only H$_2$O, CO$_2$, N$_2$, and NH$_3$ are considered, as they constitute the main products of all four compounds.
        \textbf{c},~Potential energy evolution during the coalescence of two (6,6) CNTs. 
Representative snapshots of the initial, intermediate, and final states are included in the plot. 
Gray spheres indicate carbon atoms, while purple spheres denote the atoms used for projected area calculation.
        \textbf{d},~Schematic representation of CNTs as circles. 
Purple circles denote CNTs belonging to the (6,6) family, including (6,6) itself and those derived from it, 
whereas red circles denote CNTs belonging to the (9,9) family.
The upward arrow represents coalescence, and the shading of the coalesced CNTs indicates the degree of projected area variation during the dynamic simulation. 
Coalescence of (24,24) CNTs forms an elliptical shape, indicative of their flattening.
See Supplementary Fig.~\ref{sfig:cnt} for the variation of projected area with simulation step.
        \textbf{e},~Potential energy evolution during the coalescence of two (24,24) CNTs, plotted as in \textbf{c}.
    }
    \label{fig:fig4}
\end{figure}

\begin{figure}[ht]
    \centering
    \includegraphics[width=1.0\textwidth]{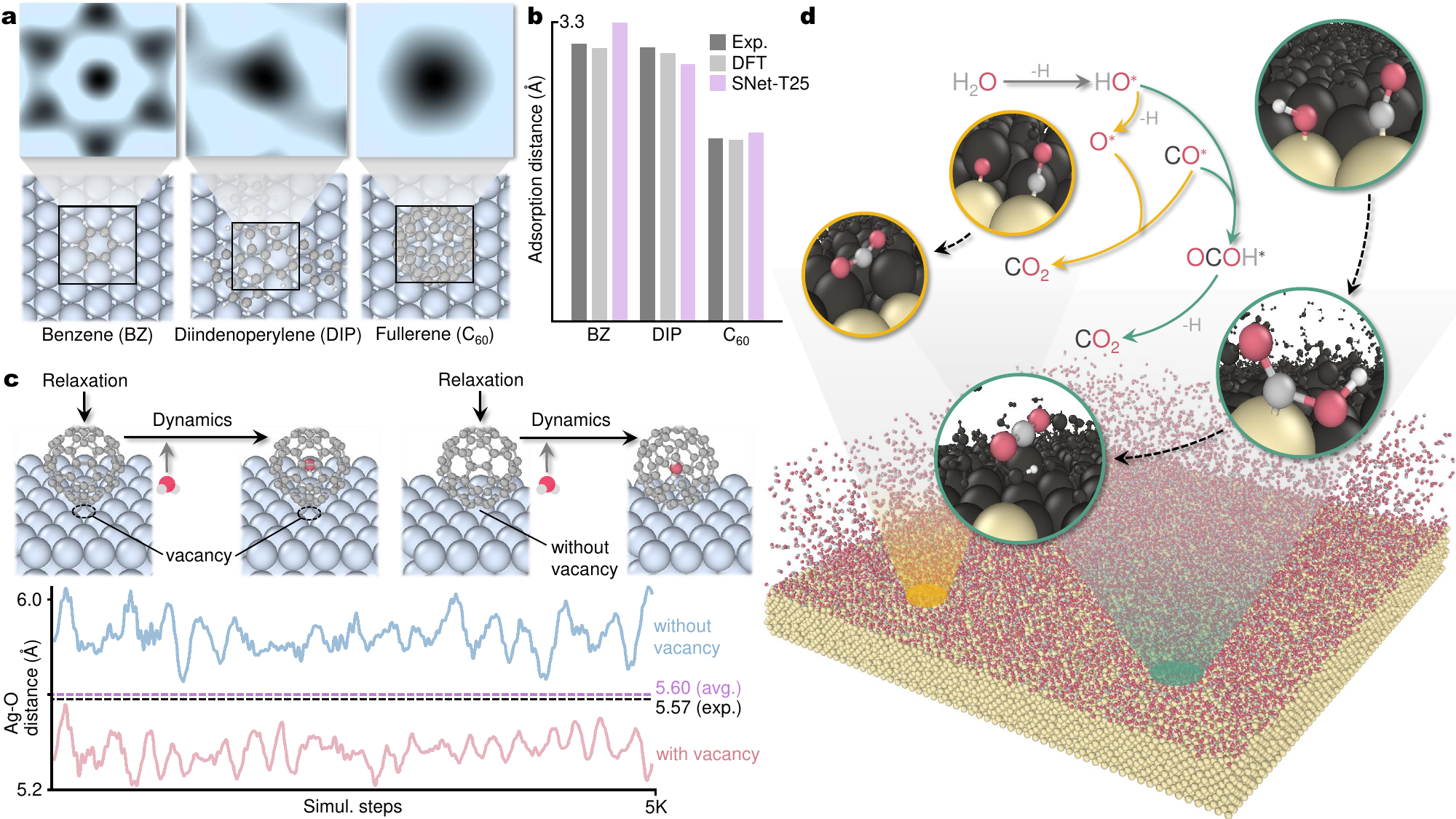}
    \caption{\small
        \textbf{MLIP-MD simulations reproduce adsorption and catalytic reaction pathways.} 
        \textbf{a},~Relaxed structures of $\pi$-conjugated molecules---BZ, DIP, and C$_{60}$---on the Ag(111) surface, 
together with their potential energy surfaces (PES) near equilibrium. 
Light-cyan, gray, and white spheres represent Ag, C, and H atoms, respectively. 
In the fullerene case, a vacancy is introduced at the center of the Ag(111) surface. 
Darker regions in the PES correspond to lower-energy configurations.
        \textbf{b},~Vertical distances of BZ, DIP, and C$_{60}$ in their relaxed structures on the Ag(111) surface. 
The vertical distance is defined as the separation between the nearest carbon atom of each molecule and the Ag(111) surface plane.
        \textbf{c},~Evolution of the Ag--O distance during MD simulations. 
The red curve denotes the case of C$_{60}$ encapsulating H$_2$O on Ag(111) with a central vacancy, 
whereas the blue curve denotes the defect-free Ag(111) surface. 
The average Ag--O distance over the two trajectories is indicated as a pale-purple dashed line (5.60\,\AA).
See Supplementary Fig.~\ref{sfig:pi-conj-md} for simulations with fixed Ag(111).
        \textbf{d},~Reaction mechanism of the water–gas shift 
obtained from MD simulations with H$_2$O and CO as reactants on an Au(100) surface terminated with OH groups. 
Ivory, gray, red, and white spheres represent Au, C, O, and H atoms, respectively.
Black spheres highlight the atoms surrounding the reactive centers for visual emphasis.
    }
    \label{fig:fig5}
\end{figure}

\clearpage


\section{Methods}\label{sec11}

\subsection{Data-generator (DG)}\label{method:dg}

The GAIA framework requires only minimal input data: 
unit-cell descriptions in POSCAR format for periodic solids or 
Cartesian coordinates in xyz format for non-periodic molecules.
Throughout this subsection, we denote these two input sets by P and NP, respectively.
Either input type can be provided independently, 
and both may be supplied simultaneously.
For the Titan25 dataset,
P includes seven metallic systems:
Ag (mp-124), Au (mp-81), Cu (mp-30), Pd (mp-2), Pt (mp-126), Rh (mp-74), and Ru (mp-33).
NP consists of isolated molecular species:
ammonia, benzene, carbon dioxide, carbon monoxide, water, C$_2$, H$_2$, N$_2$, and O$_2$.

The input structures are then processed by the builders, 
as illustrated in Supplementary Fig.~\ref{sfig:dg}, 
to generate unlabeled configurations. 
These configurations are subsequently labeled with DFT energies and forces 
obtained from single-point calculations, ab initio molecular dynamics (AIMD), 
and the initial step of the nudged elastic band method~\cite{SXCJ12}.
The remainder of this subsection details the algorithms used by each builder. 
These builders differ in how atomic species are arranged 
on the lattice or within the unit cell, 
thereby producing complementary sets of training configurations.

\bmhead{Checkerboard builder}
The Checkerboard builder generates synthetic bulk-like structures
by arranging two selected elemental species on a simple cubic lattice. 
The cubic cell has edge length $L$ 
and is filled with atoms on a uniform grid of spacing $s$, 
resulting in $(L/s)^3$ atomic sites. 
For each element pair $(A,B)$, 
stoichiometric ratios are sampled within a user-specified range 
$[r_{\min}, r_{\max}]$, 
including the symmetric 1:1 case. 
Atomic species are then distributed over the grid 
according to the chosen ratio, 
and their positions are randomly permuted 
to avoid artificial ordering. 
The cell is treated as periodic with cubic lattice vectors 
$(L,0,0)$, $(0,L,0)$, and $(0,0,L)$. 
The number of distinct stoichiometric ratios 
available for each pair is 
$r^{\mathrm{CB}}=k^2-k+1$, 
where $k=r_{\max}-r_{\min}+1$. 
For example, 
$k=1$ yields only the 1:1 ratio, 
while $k=2$ includes 1:1, 1:2, and 2:1. 
Across all unordered element pairs, 
the total number of generated checkerboard structures is 
$r^{\mathrm{CB}}\cdot\binom{|\mathrm{e}|}{2}+|\mathrm{e}|$, 
where $|\mathrm{e}|$ is the cardinality of a set of the input elements,
namely the number of available elements for both P and NP.
Minor deviations of up to one atom from the exact stoichiometry can occur 
due to integer rounding.

\bmhead{Bulk builder}
The Bulk builder generates bulk variants from the periodic inputs
by expanding each structure into a supercell,
where the replication factors $(n_a,n_b,n_c)$ are chosen 
such that the total number of atoms does not exceed the predefined threshold value (e.g., 250). 
Among all valid candidates, 
the supercell with the largest atom count is selected. 
From the resulting supercell, 
a series of perturbed structures is produced. 
First, the cell volume is modified 
by scaling factors of 1.1 and 1.2 relative to the original size. 
In addition, a fraction of atoms is randomly removed, 
with deletion ratios of 0.1 and 0.2. 
For each atom-deleted configuration, 
additional perturbations are introduced by isotropically 
scaling the volume with factors of 
0.85, 0.90, 0.95, and 1.05. 
This procedure generates 13 structures per $p\in\mathrm{P}$: 
one unperturbed supercell, 
two volume-scaled cells, 
two atom-deleted cells, 
and eight atom-deleted cells with additional volume scaling. 
In total, $13|\mathrm{P}|$ configurations are generated 
and stored for subsequent DFT labeling.

\bmhead{Slab builder}
The Slab builder shares the same supercell construction 
and perturbation procedures as the Bulk builder, 
but differs in converting the bulk supercell into a slab geometry. 
The bulk cell is cleaved along the (100) plane, 
and a vacuum gap of 30\,\AA\, is inserted 
along the third lattice vector $\mathbf{c}$ of the simulation cell. 
The structure is then translated 
so that the bottom surface lies at $z=0$. 
To maintain structural stability, 
atoms are grouped into layers by their $z$-coordinates 
within a tolerance of 0.1\,\AA, 
and 20\% of the lowest layers are constrained 
during subsequent calculations. 
This process ensures that 
the resulting slab retains in-plane periodicity 
while exposing a surface suitable for DFT labeling.

\bmhead{Adatom builder}
The Adatom builder shares the same supercell construction procedure
as the Slab builder, 
but differs by placing single-atom adsorbates 
on the slab to sample a coarse potential energy surface (PES)
above the exposed face. 
The adsorbates are defined from the unique set of elemental species 
present in the available P and NP elements. 
For each element, 
a single atom is positioned on a two-dimensional grid 
covering half of the surface unit cell, with a 1\,\AA\, margin from the edges. 
The grid is discretized into $r^{\mathrm{AD}}$ points ($5\times 5$ by default). 
At each in-plane grid point, 
the adsorbate is placed at a series of discrete heights 
$d^{\mathrm{AD}}={0.40, 0.45, 0.50, 0.55, 0.60, 0.65, 0.70, 0.80, 1.00, 1.20, 1.40, 1.60, 2.00}$\,\AA\,
above the slab, relative to the highest surface atom. 
The sampling is denser near the surface 
and becomes progressively sparser at larger distances, 
providing finer resolution for short-range interactions 
and coarser resolution for the asymptotic region.
By default, the number of configurations generated is
$N^{\mathrm{AD}}=(\binom{|\mathrm{e}|}{2} + |\mathrm{eP}| - \binom{|\mathrm{eNP}|}{2})\cdot |r^{\mathrm{AD}}|\cdot |d^{\mathrm{AD}}|$
where $|\mathrm{eP}|$ and $|\mathrm{eNP}|$ are the number of P and NP elements, respectively.
The $N^{\mathrm{AD}}$ accounts for all possible element pairs, 
adds the self-pair correction for $\mathrm{eP}$, 
and excludes pairs composed solely of $\mathrm{eNP}$.

\bmhead{Admol builder}
The Admol builder constructs adsorbed–molecule configurations 
on slab surfaces.
Conceptually, it follows the same workflow 
as the Adatom builder, 
except that the adsorbates are drawn from NP rather than single atoms. 
The molecules serve as adsorption candidates 
and are subsequently aligned, replicated, and placed above the slab.
For each pair $(p,q)\in\mathrm{P}\times\mathrm{NP}$, 
a surface supercell is prepared 
and the adsorbate $q$ is cell-synchronized to the surface reference 
so that its fractional extents fit within the surface cell.
Integer repeat numbers $(N_x,N_y)$ are chosen 
such that the in-plane dimensions of the molecular supercell match 
those of the surface cell to within 0.5\,\AA.
A rotated molecular supercell is generated 
by tiling the molecule $r^{\mathrm{AM}}$ times, 
where $r^{\mathrm{AM}} = N_x \times N_y$; 
each tile is assigned a random three-dimensional rotation 
to diversify orientation and registry. 
The tiled molecule is recentred laterally over the surface 
and placed at discrete adsorption heights $d^{AM}$ of 
1.0, 1.5, 2.0, 2.5, 3.0, 3.5, 4.0, and 4.5\,\AA\,
above the topmost surface atoms.
This placement is repeated over $|r^{\mathrm{AM}}|$ (20 by default)
independent trials 
to sample orientations and lateral registries.
By default, the number of configurations generated is
$N^{AM}=|\mathrm{NP}|\cdot |r^{\mathrm{AM}}|\cdot |d^{\mathrm{AM}}|$ 
per element in P.

\bmhead{Nanoreactor$^+$ builder}
This builder explores the reactive configurational space
of the input molecular systems 
by combining the quantum cluster growth (QCG) scheme~\cite{SPPH22} 
with the metadynamics (MTD) simulations.
This two-step procedure first generates candidate reactant assemblies 
and then drives them through reactive trajectories 
to sample bond breaking and formation events.

The QCG procedure was implemented using CREST~\cite{PBG20,SPPH22} in the QCG mode. 
Each atomic element contained in P, when available, 
was first paired with a single molecule from NP to form a pair $(p,q)\in\mathrm{P}\times\mathrm{NP}$. 
These pairs were then extended by combining the P--NP assemblies 
with additional NP partners. 
In the absence of P, 
a molecule $q$ was fixed as the solute and combined with another one from NP. 
For each pair, 
the number of partner molecules was controlled by an adjustable parameter.
Spin multiplicities were initially set to closed-shell configurations.
An initial geometry optimization could be skipped 
when the input structures were already equilibrated.
A short (1\,ps by default) MD ensemble was performed in the ALPB water model, 
while an outer wall potential confined the cluster 
with a scaling factor of 1.2 by default. 
The lowest-energy structures were collected 
and passed to subsequent Nanoreactor$^+$ simulations.
If convergence failed, 
the workflow automatically retried with progressively altered parameters: 
The wall scaling factor was varied sequentially (1.2, 1.4, 0.8, 1.0, 1.6), 
and the spin multiplicity was cycled through singlet, doublet, and triplet states. 
If all combinations failed to yield a converged ensemble, 
the corresponding pair was discarded as a failed QCG attempt.
By default, the ideal number of QCG generated configurations is
$N^{\mathrm{QCG}}= |\mathrm{P}|\cdot|\mathrm{NP}|\cdot(|\mathrm{NP}|-1) + |\mathrm{NP}|^2$.

After the QCG, the builder explores 
reactive events of the QCG-generated configurations 
by performing MTD simulations with the xTB engine~\cite{G19}.
For each input structure, 
a confining sphere was estimated from a short CREST calculation; 
when this failed, 
a random radius between 6 and 10\,\AA\, was assigned instead. 
The working radius for each trial was then drawn uniformly 
between this value and 1.5 times it, 
ensuring variability across runs.  
At each trial, the log-Fermi wall potential (6000~K) 
and biasing parameters 
($k_\mathrm{push}$, $\alpha$, and $\beta$) 
were re-sampled randomly within predefined ranges. 
Spin multiplicities were tested sequentially in a fallback manner: 
the first attempt used the doublet state, 
followed by singlet, triplet, and quartet. 
If the quartet also failed to converge, 
 the corresponding MTD simulation was discarded.
Simulations were run for 10~ps with a 1~fs integration step, 
without SHAKE constraints. 
The MTD bias was updated every 10~fs, 
and trajectories were saved every 50~fs, 
producing 200 snapshots per run. 
Each trajectory was postprocessed by converting individual snapshots 
from XYZ to SDF format with Open Babel~\cite{OBJM11}, 
and the converted structures were parsed with RDKit~\cite{rdkit}
to construct bond-order heatmaps, termed bondmaps.
The bondmaps were averaged across all snapshots within a trajectory 
to obtain a representative bond-order matrix.  
Newly generated matrices were systematically compared 
with the entire set of previously stored matrices 
using the structural similarity index (SSIM), 
and only those below a similarity threshold were retained 
as distinct products.
As a result, the total number of configurations generated by the Nanoreactor$^+$ builder is  
$N^{\mathrm{NANO}} = k^{\mathrm{NANO}} \cdot s^{\mathrm{NANO}}_{\mathrm{MTD}} \cdot N^{\mathrm{QCG}}$
where $k^{\mathrm{NANO}}$ is the user-defined number of unique bond-order matrices 
retained per configuration,  
$s^{\mathrm{NANO}}_{\mathrm{MTD}}$ is the number of snapshots 
generated in each MTD trajectory,  
and $N^{\mathrm{QCG}}$ is the number of input structures 
obtained from the QCG stage.
See Supplementary Fig.~\ref{sfig:nano} for representative configurations of 
the Nanoreactor$^+$ builder.

\bmhead{Supercell with random rotation}
The resulting Nanoreactor$^+$ configurations were combined into supercells 
before DFT labeling.  
Replication factors along the three lattice vectors were chosen 
to maximize the number of atoms
while keeping it below a user-defined limit per snapshot.
To avoid generating identical repeats, 
each replica was rotated as a rigid body 
using a rotation matrix defined by a random unit axis 
and a rotation angle uniformly sampled between 0 and 360$^{\circ}$.
Atomic coordinates were shifted to the center of mass, 
rotated, and then translated to the corresponding lattice position.  
The final simulation cell was defined with a small padding of 0.3\,\AA\, 
along each lattice direction.

\bmhead{Diatomic potential energy curves}
Potential energy curves were generated for each unique pair of elements, 
with interatomic distances sampled uniformly between user-defined minimum and maximum factors (0.6 and 3.0, respectively, for Titan25) of the sum of their covalent radii. 
The resulting energies and forces were then discretized into 30 bins and smoothed using cubic spline interpolation.

\bmhead{Dataset visualization}
Fig.~\ref{fig:fig2}b was generated using the uniform manifold approximation and projection~\cite{MHM18}.
For each data point in Titan25(G), embeddings were obtained from feature vectors extracted by the model trained on Titan25(G+I).
Each feature vector corresponds to the aggregated output of the final representation layer, 
computed from the atomic-level embeddings of the input structure.
Further details on the model architecture and training are provided in Methods~\ref{method:mt}.

\subsection{DFT labeling for Titan25}\label{method:dl}

All DFT calculations were performed with VASP.6.4.1~\cite{KF96a,KJ99}.
Projector-augmented wave (PAW) pseudopotentials 
and the PBE functional were used. 
A plane-wave cutoff energy of 520~eV was applied, 
and van der Waals interactions were included 
through the D3 correction with the zero-damping function.
Brillouin-zone sampling was restricted to the $\Gamma$ point. 
Electronic convergence was set to 10$^{-5}$~eV, 
and atomic forces were converged to below 10$^{-4}$~eV/\,\AA. 
Partial occupancies were treated using Gaussian smearing
with a width of 0.05~eV.  
Symmetry constraints were turned off.
Configurations generated by the Adatom and Admol builders 
were evaluated by single-point energy calculations 
under these conditions.  

\bmhead{Structure relaxation}
For the configurations generated by the Checkerboard builder, 
structural relaxations were performed using the conjugate-gradient algorithm 
with a relaxation step size of 0.2 for up to 20 ionic steps.  
During these relaxations, only the atomic positions were optimized 
while the cell shape and volume were kept fixed. 
Whether converged or not, the resulting structures were used as initial conditions 
for subsequent AIMD simulations.  
These relaxations served only to generate suitable starting configurations for AIMD, 
rather than to obtain fully optimized structures.  

\bmhead{Ab initio MD}
Two AIMD protocols were used to generate DFT-labeled datasets,  
with each protocol applied to configurations generated by different builders.  
The first protocol was carried out in the NVT ensemble using the Nose-Hoover thermostat  
for 1,000 steps with a timestep of 0.5~fs,  
during which the temperature was increased from 400~K to 600~K.  
This protocol was applied to datasets generated by the Checkerboard, Bulk, and Slab builders.  
The second protocol employed the same simulation parameters as the first,  
but the trajectories were limited to 20 steps.  
This shorter protocol was applied to Nanoreactor$^+$ configurations.  
In both protocols, a fictitious Nose-Hoover mass parameter of 20 was used.  

\bmhead{NEB initial step}
Trajectories generated by the Nanoreactor$^+$ builder 
were further used to construct connected structures 
via the nudged elastic band scheme.
Representative snapshots were taken at regular intervals, 
and initial and final configurations were paired to produce intermediate images 
through linear interpolation using the \texttt{nebmake.pl} utility provided with VASP.
These interpolated configurations were not subjected to full NEB optimizations;  
instead, they were directly evaluated by single-point DFT calculations 
to provide reference energies and forces.  
Before the DFT labeling,
all generated structures were screened for unphysical short contacts.  
A~Coulomb-like repulsion estimate was used, defined as
$E_\mathrm{rep} = k/r$,
where $k=14.4~\mathrm{eV}\!\!\cdot\!\!\text{\AA}$ 
and \(r\) is the shortest interatomic distance.  
Configurations with $E_{\mathrm{rep}} > 50~\mathrm{eV}$ were discarded,
and only the remaining physically reasonable structures were retained.  
As described in Methods~\ref{method:dg}, the retained configurations were further expanded into supercells 
with randomized rigid-body rotations to enhance structural diversity.  

\subsection{Data-improver (DI)}\label{method:di}

As outlined in Fig.~\ref{fig:fig2}c,
DI serves to analyze the deficiencies in the exploratory data from DG and to supplement them accordingly.
Its detailed procedure is described as follows.

\bmhead{Categorical analysis}
As the first stage of DI, 
categorical analysis is performed on the generated data points 
to explore undersampled areas in chemical space,
which can be low-density regions or error-prone domains.
The key here lies in establishing criteria for data categorization;
we propose a two-dimensional scheme for this purpose.

In the scheme, one axis corresponds to the conventional measure of total energy,
the other to a diversity metric $d_\Lambda$,
given by the standard deviation of eigenvalues of interatomic distances in a snapshot.
Specifically, it is obtained as
$d_\Lambda(i) = ( \sum_{j=1}^{n_i} (\lambda_j - \bar{\lambda})^2 / n_i )^{1/2}$,
where $n_i$ is the number of atoms in snapshot $i$, 
and $\lambda_j$ is the $j^{\text{th}}$ eigenvalue of the atom--atom distance matrix.
The proposed metric thus indicates how much configurational diversity is present in the snapshot.

We note that variations in the number of atoms across data points were not explicitly handled 
in the measurement of the standard deviation.
This could introduce biases in purely geometric comparisons,
since the corresponding normalization was not applied.
Nevertheless, such normalization could, in turn, suppress genuine contributions 
from supercell enlargement, such as inter-complex interactions with random orientations, 
which are essential sources of structural diversity. 
Hence, the design choice of omitting normalization reflects our focus on 
capturing not only shape-based differences 
but also the overall interaction modes relevant for MLIP training.
Yet, to mitigate the possible geometric mismatch arising from variations in atom count,
the maximum system size in a data point was restricted to 250 atoms.
With this perspective, leaving the raw spectra allows accounting for
the comprehensive modes in atomic environments represented across snapshots,
thereby enabling the remaining stages to cover as broad a structural variety as possible.

With the user-defined values $M_E$ and $M_\Lambda$,
the two-dimensional space is divided into an $M_E \times M_\Lambda$ grid,
where each cell of this grid stores (i) the number of corresponding data points from the training dataset 
and (ii) the ratio of outliers in the validation dataset within that cell.
Outliers refer to data points with prediction errors exceeding the user-defined threshold.
In our case, we used force RMSE as the error metric and set the threshold to 2.0.
Accordingly, cells with lower values for the training dataset and higher values for the validation dataset 
suggest the presence of sparse (or absent) coverage across both observable and hidden regions.

\bmhead{Basis snapshots sampling}
Guided by the improvable coverage above, basis snapshots are data points from which the new seed structures are selected to further span the current chemical space.
To do this, we construct the probability distributions for the sampling,
defined by 
$P_F(x_{ij}) = (\sum_{i',j'} x_{ij} / x_{i'j'})^{-1}$ and
$P_P(\xi_{ij}) = \xi_{ij} / \sum_{i',j'} \xi_{i'j'}$
for the frequency-based sampling on the training dataset and the prediction-based sampling on the validation dataset, respectively,
where $x_{ij}$ is the number of data points in the cell $C_{ij}$, and
$\xi_{ij}$ is the ratio of outliers therein.
In the current implementation, the number of basis snapshots is user-specified 
for both sampling strategies.

\bmhead{Seed expansion}
Serving as the core of the seed expansion,
the extraction of chemical substructures is performed on the selected basis snapshots.
First, the positional coordinates of a snapshot are unwrapped in order to restore continuity across periodic boundaries.
By creating an imaginary cube with side length $2{\delta}r_{\text{COV}}$ around each atom, 
where $r_{\text{COV}}$ is the covalent radius and $\delta$ is the volume tolerance,
bond presences are identified by the overlap of the cube pairs.
Accordingly, chemical substructures in each snapshot are then extracted by the depth-first search.
Among them, those with less than a specific number of atoms---the minimum value 
within the Nanoreactor$^+$ results from the previous DG module---are filtered out.
Finally, new seeds are determined from the extracted substructures 
in consideration of the previous $N_\text{NANO}$,
which are fed back into the Nanoreactor$^+$ to further generate the exploratory data points.

\subsection{Model training}\label{method:mt}
To train MLIPs on Titan25 and other datasets for comparison,
we used an equivariant graph neural network---specifically SevenNet~\cite{PKHH24m}, 
a variant of NequIP~\cite{BMSG22} in which the self-connection parameters are shared across all chemical elements.
The model architecture was configured with 
an embedding block having a 128$\times$0e output dimension 
and three interaction blocks, 
where the first two have output features of 128$\times$0e+64$\times$1e+32$\times$2e+32$\times$3e 
and the last has 128$\times$0e, followed by the readout block.
The training was performed using an in-house framework built on top of FAIRChem~\cite{fairchem},
extended to support SevenNet and 
to enable the fused kernels such as cuEquivariance~\cite{cueq} and FlashTP~\cite{LKPJ25}.
For a fair comparison across datasets, 
the number of training steps was fixed at approximately 0.5 million with the same batch size. 
Accordingly, depending on dataset size, the model was trained for 100, 125, and 80 epochs 
on MPTrj, Titan25(G) and Titan25(G+I), respectively.
In the case of ANI-1xnr, which has a much smaller dataset size than the others, 
there could be a risk of overfitting with overly prolonged training; thus, the training was limited to 300 epochs.
During dataset preparation, 
since the test datasets in GAIA-Bench were constructed using PBE-D3,
ANI-1xnr was re-labeled with the PBE-D3 functional under the same conditions as in Methods~\ref{method:dl}.
For MPTrj, D3 correction terms were calculated independently and added for a consistent comparison.
The data in each dataset were split into training, validation, and test sets with an 8:1:1 ratio.
Further training details are provided in Supplementary Table~\ref{stab:model_training}; 
the validation results on Titan25(G+I) are included 
in Supplementary Figs.~\ref{sfig:loss}~and~\ref{sfig:parity_valid}.

\subsection{GAIA-Bench}\label{method:gb}

GAIA-Bench comprises four benchmark tasks---mol2mol, bulk, slab, and mol2surf---with 
reference data obtained from DFT calculations. 
We used the same DFT parameters and conditions as in Titan25. 
In this subsection, we describe the implementation details of GAIA-Bench.

\bmhead{Structure perturbation}
For force evaluations, 
perturbed configurations were generated by applying randomized displacements 
to the same structures used for energy comparisons in each benchmark. 
Each perturbed structure was obtained through multiple trial displacements, 
with only those meeting the acceptance criteria retained. 
At each trial, atoms were randomly displaced 
by 0.01--0.20\,\AA\, (0.01--0.1\,\AA\, for mol2mol) 
along directions biased by local neighbor vectors, 
with small random angular deviations added to the displacement directions.  
If the target force distribution was not sufficiently sampled after 10 iterations, 
and the maximum atomic force remained below 1.2 times the target, 
the upper displacement bound was gradually increased by 1\% per step 
until a maximum of 1.0\,\AA\, was reached.  
Perturbed structures were accepted 
only if at least 30\% (20\% for mol2mol) of atoms 
had forces within the target range, 
while the maximum force remained below 1.2 times the target.  
For mol2mol, target forces of 0.1, 1.0, 3.0, and 5.0~eV/\,\AA\, were considered,  
whereas bulk and slab employed targets of 3.0, 5.0, and 7.0~eV/\,\AA. 
A total of 3,024, 4,713, 870, and 80,424 configurations were generated 
for force calculations in mol2mol, bulk, slab, and mol2surf, respectively.
The axis-wise force MAE values were then evaluated against the DFT forces,
with the analysis performed for each group 
as defined within the individual benchmark tasks.  

\bmhead{mol2mol}
The molecular interactions of a set of nine molecules were investigated: 
1,2,4-trimethylbenzene, 
1,2-ethanediol, 
1-butanol, 
1-propanol, 
2-butanol, 
1,4-dioxane, 
nitroethane, 
propylamine, 
and water, 
consisting solely of C, H, N, and O. 
Structural information for these molecules 
was retrieved from PubChem~\cite{KCCG25}.
The near-equilibrium structures of molecular pairs, 
including self-pairs,
were identified using the QCG scheme with the GFN2-xTB method~\cite{SPPH22}.
From those near-equilibrium structures,  
one molecule was rotated around a random axis 
to generate 30 distinct orientations, 
sampling diverse molecular interactions.
From each rotated molecular pair,  
the interatomic vector connecting the closest atoms between the two molecules 
was determined and used as the reference axis.
Along this axis, 
the intermolecular separation was scaled 
from 0.5 to 3.5 times the original distance.
The resulting 22,650 configurations 
were classified into 81 distinct groups
based on the molecular components of the pairs.
For each group,
the relative energies were evaluated with respect to the configuration
having the largest intermolecular separation,
enabling the calculation of interaction energies.

\bmhead{bulk}
Periodic bulk structures were obtained from the Materials Project~\cite{JOHC13},  
considering assemblies of Ag, Au, Cu, Pd, Pt, Rh, Ru, O, and H.  
All unary, binary, and ternary combinations were included,  
except for sets containing only O and H.  
Supercells were constructed by repeating the conventional cell 
to maximize the number of atoms under a cap of 200.  
To sample volumetric strain states, 
the cells were isotropically scaled with factors of 
0.85, 0.90, 0.94, 0.98, 1.00, 1.02, 1.04, 1.06, 1.08, 1.10, 1.12, 1.14, and 1.16, 
with atomic positions scaled accordingly, 
yielding a family of uniformly strained supercells for each composition.  
In total, 1,783 configurations were generated by this procedure.  

To construct energy--volume curves, 
structures sharing the same composition were grouped.  
DFT calculations that did not converge or failed were removed,  
which sometimes reduced the group size.  
Only groups retaining at least two valid structures were used in the analysis,  
resulting in 143 valid groups out of 144.  
For each group, the structure with the lowest DFT total energy was identified  
and designated as the reference geometry.  
This geometry was then re-evaluated with the MLIP,  
and relative energies were computed 
by subtracting the corresponding reference energy in each method.  
This procedure ensured a consistent baseline 
while allowing direct comparison between DFT and an MLIP.  

\bmhead{slab}
The structures for the slab benchmark were
obtained similarly to the bulk benchmark,
but only binary and ternary combinations were considered.
Surface indices (001), (110), and (111) were generated 
with two layers and a vacuum spacing of 30\,\AA,
ensuring that the total number of atoms did not exceed 400.
A total of 392 configurations were generated for the slab benchmark.

With reference to the (110) surface,
the relative energies of (001) and (111) surfaces
for the same components were considered as a group.
If the DFT calculation for the (110) surface failed for a group,
the (111) surface was used instead.
A total of 167 groups were categorized.

\bmhead{mol2surf}
The mol2surf benchmark comprises 
configurations of molecular bilayers on periodic surfaces.
These surface structures were generated using the same methodology 
as the slab benchmark,
with molecules taken from the mol2mol benchmark set.
In contrast to the slab benchmark,
the mol2surf benchmark considered only the (001) and (111) surface indices,
with a maximum of 200 atoms in the supercell.
For each surface,
two molecular pairs were chosen
and arranged in a bilayer configuration with a 1.5\,\AA\, offset.
The molecules were then replicated with random rotations,
following the DG procedure.
For each configuration,
three random rotations of the bilayer molecules were performed,
and the resulting configurations were grouped together.
A total of 80,424 configurations were generated, 
resulting in 26,908 groups.
The energy of each group was assessed relative to its first configuration.

\subsection{MLIP-MD simulations}\label{method:MD}

MD simulations were conducted primarily using LAMMPS version 2 Aug 2023~\cite{TABB22},
employing the Titan25-trained model SNet-T25,
whose checkpoint was converted for domain-decomposed parallel MD.

\bmhead{Pt-CO$_x$-H$_2$O}\label{method:ptcox}
To investigate CO or CO$_2$ on the Pt(111) surface 
in an explicitly solvated environment,
MD simulations were performed.
The initial cubic Pt unit cell structure 
was taken from the Materials Project (ID: mp-126).
Supercells with fewer than 500 atoms were constructed 
to create four-layer Pt(111) surfaces for both CO and CO$_2$ cases.
For CO, 96 molecules formed a double layer,
overlaid by 256 water molecules in four layers.
For CO$_2$, 36 molecules were placed on the surface in a double layer,
followed by 125 water molecules in five layers.

Both simulations were conducted 
using the Nose-Hoover thermostat at 300~K.
The simulations consisted of 50,000 steps with a 0.5~fs timestep,
resulting in a total simulation time of 25~ps.
Snapshots were recorded every 100 steps
to analyze the vertical distance between the plane 
constructed from the atoms in the outermost Pt surface layer and the carbon atoms.
For each snapshot, we computed the perpendicular distance between this reference plane 
and all carbon atoms. 
A global cutoff (1.5\,\AA\ for CO and 4.0\,\AA\ for CO$_2$) was applied, 
such that only carbon atoms located within those distances from the Pt plane 
were included in the analysis. 
The reported values correspond to the mean perpendicular distance of these atoms 
at each snapshot, 
and trajectories in which no carbon atom satisfied the cutoff 
were excluded from the averaging procedure.

For both systems,
simplified structures were used for additional DFT calculations.
The four-layer Pt(111) surface, comprising 96 atoms,
was used to place 6 CO (or CO$_2$) molecules in a double layer,
accompanied by 24 water molecules in four layers.
The simulations employed the PBE-D3 functional 
and Nose-Hoover thermostat at 300~K,
consisting of 20,000 steps with a 1.0~fs timestep,
resulting in a total simulation time of 20~ps.

Using these simulations,
we compared SNet-T25 against UMA-M~\cite{WDFG25m}.
The UMA-M model comprises approximately 1.4 billion parameters in total,
of which about 50 million are active for a given atomic structure,
about 100 times the total parameter count of SNet-T25 (0.5 million).
It was trained on a combined dataset spanning 
OC20~\cite{CDGL20}, OMat24~\cite{BSFW24m}, OMol25~\cite{LSST25m}, ODAC25~\cite{SBYC25}, and OMC25~\cite{GBYS25}.
In UMA-M, a single set of model parameters is shared across the five datasets, 
and the forward path is adapted to the input task during inference through task-dependent routing.
We additionally note that,
while simulations in this subsection were carried out mainly using LAMMPS (as stated above), 
for the Pt-CO$_x$-H$_2$O systems we performed simulations of both UMA and Titan25-trained models using ASE v3.25.0~\cite{LMBC17} to ensure consistency,
at a time when the FAIRChem-LAMMPS~\cite{fairchem} interface for the official UMA checkpoints~\cite{uma} was not yet available.

\bmhead{Hugoniostat simulation of energetic molecules}
The isotropic Hugoniostat simulations were conducted on four energetic molecules:
RDX, TNT, TATB, and nitromethane,
whose molecular structures were obtained from PubChem.
For each system, one thousand molecules were placed in a cubic cell  
with a 4\,\AA\, offset applied to each molecule. 

To identify products,
molecules in MD trajectory files were analyzed
using a graph-based approach where atoms were nodes
and bonds were edges based on interatomic distances.
Bonding was determined by checking if the distance between two atoms
was within 0.3 to 1.5 times the sum of their covalent radii.
Each resulting connected component represented a distinct molecular species.
A molecule was identified as H$_2$O
if the component consisted of one oxygen and two hydrogen atoms 
and the H--O--H angle fell within the range of 
73--135.5$^{\circ}$ (0.7--1.3 $\times$ 104.5$^{\circ}$).
For CO$_2$, a component was counted as CO$_2$ 
if it consisted of one carbon and two oxygen atoms 
and the O--C--O angle was within 126--180$^{\circ}$ (0.7--1.0 $\times$ 180$^{\circ}$).
A molecular fragment was classified as NH$_3$ 
if it contained one nitrogen and three hydrogens,  
and all H--N--H bond angles lay within 30~\% 
of the ideal tetrahedral angle (107$^{\circ}$). 
For each species, the normalized molecular counts were averaged
over the last ten snapshots of the trajectory.

\bmhead{CNT simulation}
CNT structures were generated using TubeASP~\cite{tubeasp},
with an initial C--C bond length of 1.42\,\AA.
For each chiral index, the corresponding structure consisted of 12 unit cells.
Initial configurations were prepared for pairs of CNTs
with identical chiral indices,
and the inter-CNT distance was set to 3.0\,\AA.

The coalescence of CNTs was simulated
using a hybrid approach combining 
kinetic Monte Carlo (kMC) and MD methods,
as implemented in the Papreca package~\cite{NEDT24}. 
At the contact interface between the two CNTs,
the simulation protocol involved severing the interfacial C--C bonds 
and subsequently performing 100 steps of NVT MD simulations at 1000~K
with a timestep of 0.5~fs, 
without any additional bias on the chemical bonds.
Here we note that
since the kMC event rates are not calibrated to physical kinetics,
the reported time should not be interpreted as a physical timescale.

To analyze the cross-sectional geometry of carbon nanotubes,
the atomic coordinates were projected onto a plane normal to the tube axis.
DBSCAN clustering~\cite{EKSX96} was applied to the atomic positions in each frame
to identify individual tube cross-sections.
Ellipse fitting using PCA was performed on the projected coordinates
to determine the semi-major and semi-minor axes.
The cross-sectional area was then calculated using 
$\pi a b$, where $a$ and $b$ denote the length of fitted semi-axes.

\bmhead{$\pi$-conjugated molecules on metallic surfaces}
The unit cell structures for face-centered cubic metals (Ag, Au, Cu, Pt)
were retrieved from the Materials Project.
Using these unit cells, 
four-layer (111) surfaces with 256 atoms were constructed
with a 30\,\AA\, vacuum layer.
The molecular structures of benzene and diindenoperylene 
were retrieved from PubChem,  
and that of fullerene from ref.~\cite{fullerene}.
For the initial relaxation, 
the adsorbate was positioned 2.0\,\AA\, above the surface.
For the fullerene system,
a modified Ag(111) surface with a central vacancy
was employed to replicate the experimental conditions.
For H$_2$O encapsulated in the fullerene,
the molecule was centered within the fullerene cage
and subsequently placed on the Ag(111) surface;
configurations both with and without a central vacancy were considered.

For both SNet-T25 and UMA-M models,
relaxations were carried out using the BFGS algorithm in ASE.  
Atomic positions were relaxed 
until the maximum residual force fell below 0.001~eV/\,\AA.
MD simulations were conducted
to determine the vertical distances between the Ag(111) surface 
(with and without vacancy)
and the O atom of the H$_2$O molecule encapsulated within fullerene.
The simulations employed the Nose-Hoover thermostat at 180~K
and consisted of 50,000 steps with a 0.5~fs timestep,
yielding a total simulation time of 25~ps.

\bmhead{Water-gas shift reaction}
The unit cell structure of Au was retrieved 
from the Materials Project (ID: mp-81).
A five-layer Au(100) surface was then generated
by replicating the unit cell,
yielding a total of 69,620 Au atoms.
The Au surface was then randomly terminated with 4,462 OH molecules.
After OH termination, 
4,332 CO molecules were deposited in three layers 
(1,444 molecules per layer), 
followed by three additional layers of H$_2$O 
comprising 4,557 molecules in total (1,519 per layer),
with a 30\,\AA\, vacuum layer.

MD simulations were performed using the Nose-Hoover thermostat 
at a temperature of 600~K.
The bottom layer of Au atoms was fixed, 
and a reflecting wall was placed 
at 80\,\AA\, from the bottom of the simulation box 
to confine the system along the z-direction.
The simulation consisted of 200,000 steps
with a 0.5~fs timestep.

\subsection{Data availability}\label{method:data_avail}
The Titan25 dataset is available on Hugging Face at \url{https://huggingface.co/datasets/aixsim/Titan25}.
The GAIA-Bench snapshots are available at \url{https://huggingface.co/datasets/aixsim/GAIA-Bench}.

\subsection{Code availability}\label{method:code_avail}
The GAIA source code and the GAIA-Bench evaluation code are available on GitHub at \url{https://github.com/samsungDS-PoCs/GAIA}.
The SNet-T25 model checkpoint is available at \url{https://huggingface.co/aixsim/SNet-T25}.


\backmatter



\clearpage


\setcounter{figure}{0}
\renewcommand{\figurename}{Supplementary Fig.}
\renewcommand{\thefigure}{\arabic{figure}}
\setcounter{table}{0}
\renewcommand{\tablename}{Supplementary Table}
\renewcommand{\thetable}{\arabic{table}}
\section{Supplementary information}

\subsection{Related work}\label{supp:related_work}

In this subsection, we review prior work and contextualize the contributions of GAIA.

\bmhead{MLIP datasets}
Since the Titan25 dataset has been developed for general-purpose use,
here we focus primarily on the corresponding datasets, rather than system or task-specific ones.

Following the advent of Materials Project~\cite{JOHC13s},
numerous datasets have been introduced for bulk crystal structures.
MPF~\cite{CO22} and MPTrj~\cite{DZJR23s} were pioneering efforts that curated
data from the Materials Project into MLIP datasets
at the million scale.
Subsequently, 
dataset sizes have progressively grown, 
through GNoME~\cite{MBSA23}, MatterSim~\cite{YHZL24}, and Alexandria~\cite{SCRL24},
culminating in the OMat24 dataset \cite{BSFW24s} that contains 100 million data points.

Moving beyond bulk structures,
OC20~\cite{CDGL20s} and OC22~\cite{TLSW22} addressed the adsorption of CHNO-based adsorbates on slabs,
thereby capturing interfacial interactions at surfaces,
which are also among the central goals of our approach.
ODAC23~\cite{SCYB24} and ODAC25~\cite{SBYC25s} focused on a different type of adsorption, 
namely the adsorption of CO$_2$, H$_2$O, N$_2$, and O$_2$ on metal-organic frameworks.
In addition, the molecular datasets such as OMol25~\cite{LSST25s} and OMC25~\cite{GBYS25s} were developed very recently.
OMol25 provides extensive coverage of intra- and intermolecular interactions,
including diverse charge and spin states, solvation, and reactivity.
OMC25 was introduced as a molecular crystal dataset.

In the context of chemical reactivity,
the Transition1x~\cite{SBVB22} 
and Halo8~\cite{LJAU25}
datasets incorporated reactivity by leveraging 
the nudged elastic band~\cite{SXCJ12s}
and single-ended growing string method~\cite{Z15}
to obtain transition states connecting reactants and products.
OMol25 has sought to achieve this through geodesic interpolation~\cite{ZTM19} 
and sampling based on the artificial force induced reaction~\cite{SMM16}, 
which enables the identification of numerous reaction pathways by 
applying artificial forces to specific atom pairs (or fragments) to induce bond rearrangements.
However, the targeted nature of the search can make it challenging for these approaches
to perform comprehensive global exploration.
To this end, Nanoreactor~\cite{WTML14} offers a promising alternative,
providing efficient discovery of reactive events with minimal a~priori mechanistic assumptions 
under high-temperature conditions.
Using the Nanoreactor, ANI-1xnr~\cite{ZMJK24s} was developed 
as a pioneering dataset for general condensed-phase reactions involving C, H, N, and O.

Beyond the prior efforts,
the proposed GAIA, together with its resulting dataset Titan25, is characterized by 
three distinctive features:
\begin{itemize}
\item GAIA facilitates de novo design of exploratory structures,
allowing open-ended generation of varied snapshots.
By contrast, most of the existing datasets partially or fully rely on reference databases for structure construction.
For instance, the crystalline datasets and the bulk structures in OC20 and OC22 
were based on the Materials Project database,
the metal-organic framework datasets on CoRE MOF 2019~\cite{CHBH19},
OMC25 on the OE62 dataset~\cite{SKGT20},
and OMol25 on multiple databases.

\item GAIA can provide effective diversity to represent a wide range of chemical reactions.
The diversity is established using a Nanoreactor implementation following ref.~\cite{G19s},
where the expensive ab initio treatment is replaced by
a semi-empirical tight-binding method with metadynamics.
Compared with the MLIP-driven Nanoreactor in ANI-1xnr,
the design of GAIA circumvents the need of iterative MLIP training 
to drive dynamics-based exploration.

Furthermore, as an extension of Nanoreactor,
we introduced three additional treatments to Nanoreactor$^+$
to achieve better exploration:
(i) Quantum cluster growth~\cite{SPPH22s} is adopted for the initial mixture arrangement
of the user-specified input components,
rather than relying on manual or random placement.
The resulting arrangements serve as the initial configuration for exploration.
(ii) The outputs of Nanoreactor are filtered
according to the bondmap similarity criterion,
discarding results similar to previously generated ones and
collecting until the desired number is reached.
(iii) Random rotations are applied during the repeated placement 
of the generated structures to construct supercells,
thereby better capturing the interactions involved.

\item Titan25 enables MLIPs to achieve improved performance on interfacial reactions.
This is accomplished by the exploratory builder, Nanoreactor$^+$,
which produces diverse metal/metalloid–nonmetal interactions, 
alongside five determined structure builders that handle rule-based generation.
In our evaluation, 
the Titan25-trained model achieved improved results on all the four GAIA-Bench tasks 
compared with the models trained on ANI-1xnr and MPTrj (Fig.~\ref{fig:fig3}a--c).
It also maintained better performance in actual simulations,
outperforming the UMA model~\cite{WDFG25s} trained on multiple foundation datasets 
in a unified manner (Fig.~\ref{fig:fig3}d,e).
\end{itemize}

\bmhead{ML-guided structure augmentation}
In one line of research, there have been active learning (AL) studies, including ANI-1xnr (mentioned above),
that leverage the concept of uncertainty for the structure augmentation.
The uncertainty here is usually defined as 
the degree of disagreement among an ensemble of models
(e.g., the query-by-committee~\cite{SOS92}),
typically quantified by the variance of their predictions.
In UDD-AL~\cite{KBLL23s} and HAL~\cite{OSKO23s},
the uncertainty estimation is applied in two ways:
(i) as a bias in the dynamics so that the sampling is performed toward regions of high uncertainty, and
(ii) as a criterion to prioritize structures for quantum-mechanical labeling,
thereby promoting an efficient expansion of the accessible chemical space.
As an alternative to the dynamics-based strategies,
there also exist AL approaches that
employ static optimization methods, such as NEB and SE-GSM~\cite{Z15},
in the exploration of reaction pathways~\cite{ASMB25, KCDM25}.
In this regard, a recent work, ArcaNN~\cite{DPGA25}, provides a structural framework
for the automated implementation of such uncertainty-based methodologies.

However, these methods face two notable limitations:
First, 
the AL methods require a priori data points---both  
as a training dataset to build an initial MLIP 
(if a pretrained model is available, initial MLIP training can be skipped)
and as a candidate reservoir of initial configurations for structure exploration.
Instead,
the generator-improver design of GAIA enables the data-improver's structure augmentation stage
to operate directly on new substructures extracted from data points produced by the data-generator,
thereby facilitating exploration with new seed components materially distinct from those originally provided by the user.
Second, 
the AL methods introduce additional computational overhead
for training multiple models to build an ensemble across AL iterations.
That is, if the ensemble consists of $N$ (e.g. 8) MLIPs, each of them needs to be trained separately.
Although several studies have employed a concept of evidential learning~\cite{SKK18, ASSR20}
to estimate uncertainty with only a single model~\cite{SAGR21, XCTM24},
the cost of model training remains substantial.
By contrast, the data-improver of GAIA operates without additional overhead beyond the structure search process;
it only demands the model inference cost on relatively small validation data
for the prediction-based sampling and the extraction cost of chemical substructures,
which is substantially lower than that of model training.
In this context, several works have attempted to avoid reliance on pre-existing models in structure search~\cite{BCD19s, LMEF25s}, similar to our approach. 
However, their exploration typically starts from broadly random sampling of candidate structures,
which can demand extensive sampling to adequately cover diverse regions of chemical space.

\clearpage
\subsection{GAIA and Titan25}\label{supp:gaia_titan25}

\begin{figure}[H]
    \centering
    \includegraphics[height=0.65\textheight, width=1.0\textwidth, keepaspectratio]{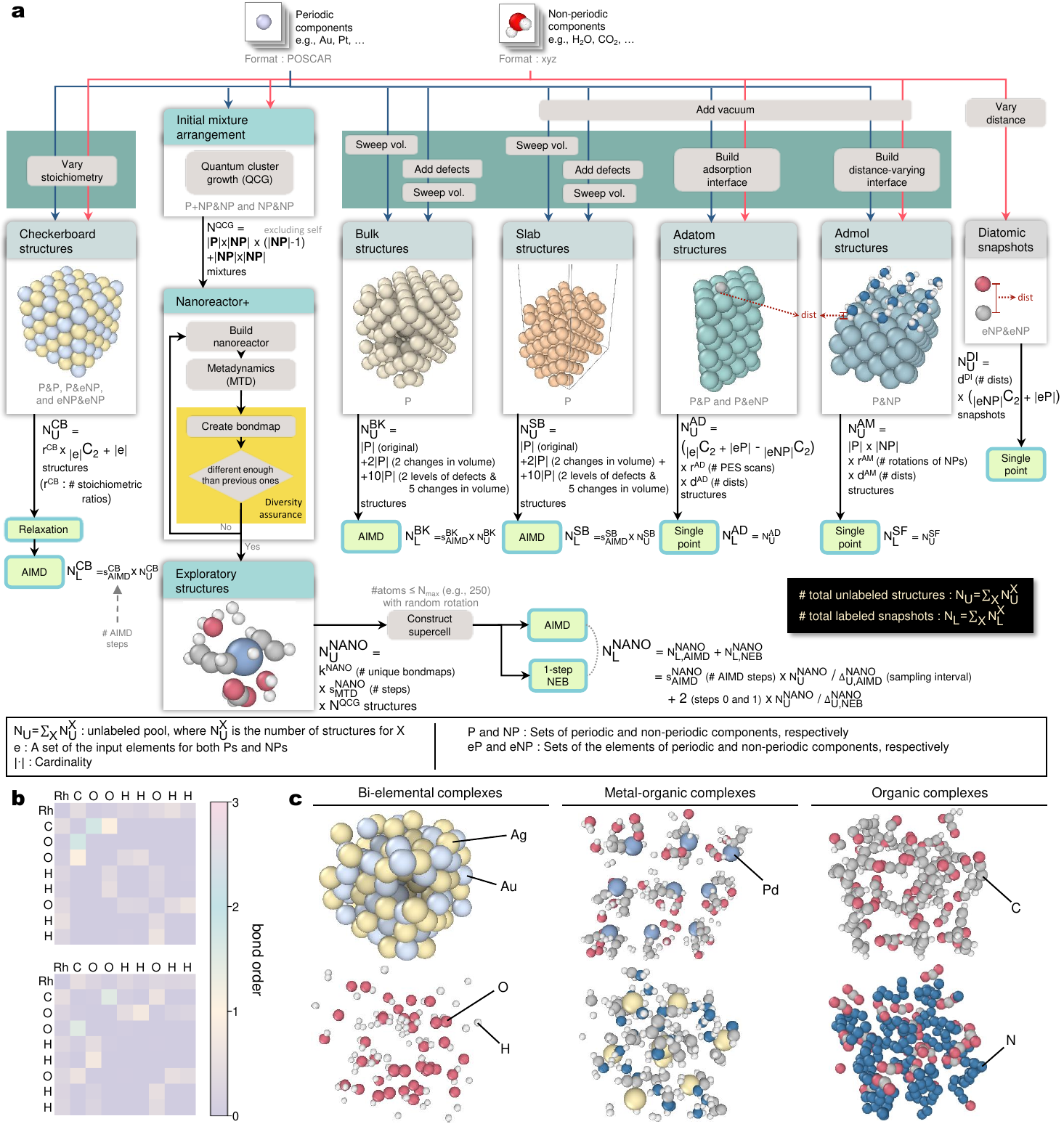}
    \caption{\small
        \textbf{a},~Details of GAIA data-generator. 
Various combinations of atomic arrangements are generated using the user-provided inputs (top) and the builders described below. 
$N^{\text{CB}}_\text{U}$, $N^{\text{NANO}}_\text{U}$, $N^{\text{BK}}_\text{U}$, $N^{\text{SB}}_\text{U}$, $N^{\text{AD}}_\text{U}$, and $N^{\text{AM}}_\text{U}$ 
indicate the maximum number of unlabeled structures that can be generated by each builder (Checkerboard, Nanoreactor$^+$, Bulk, Slab, Adatom and Admol) 
and $N^{\text{DI}}_\text{U}$ for diatomic curve add-on. 
The same symbols with the subscript ``L'' denote the maximum number of labeled structures. 
For detailed explanation, see Methods~\ref{method:dg}.
        \textbf{b},~Illustration of bondmap examples. 
Averaged bond matrices of trajectories generated from a mixed structure comprising a Rh single atom, a single CO molecule, and two H$_2$O molecules are shown. 
The left and right panels depict the first and second unique averaged bond matrices derived from this mixed structure, obtained with an SSIM threshold of 0.98, respectively.
        \textbf{c},~Representative configurations generated by the Checkerboard (bi-elemental complexes) and the Nanoreactor$^+$ builder (metal–organic and organic complexes).
    }
    \label{sfig:dg}
\end{figure}

\clearpage
\begin{figure}[ht]
    \centering
    \includegraphics[width=0.9\textwidth]{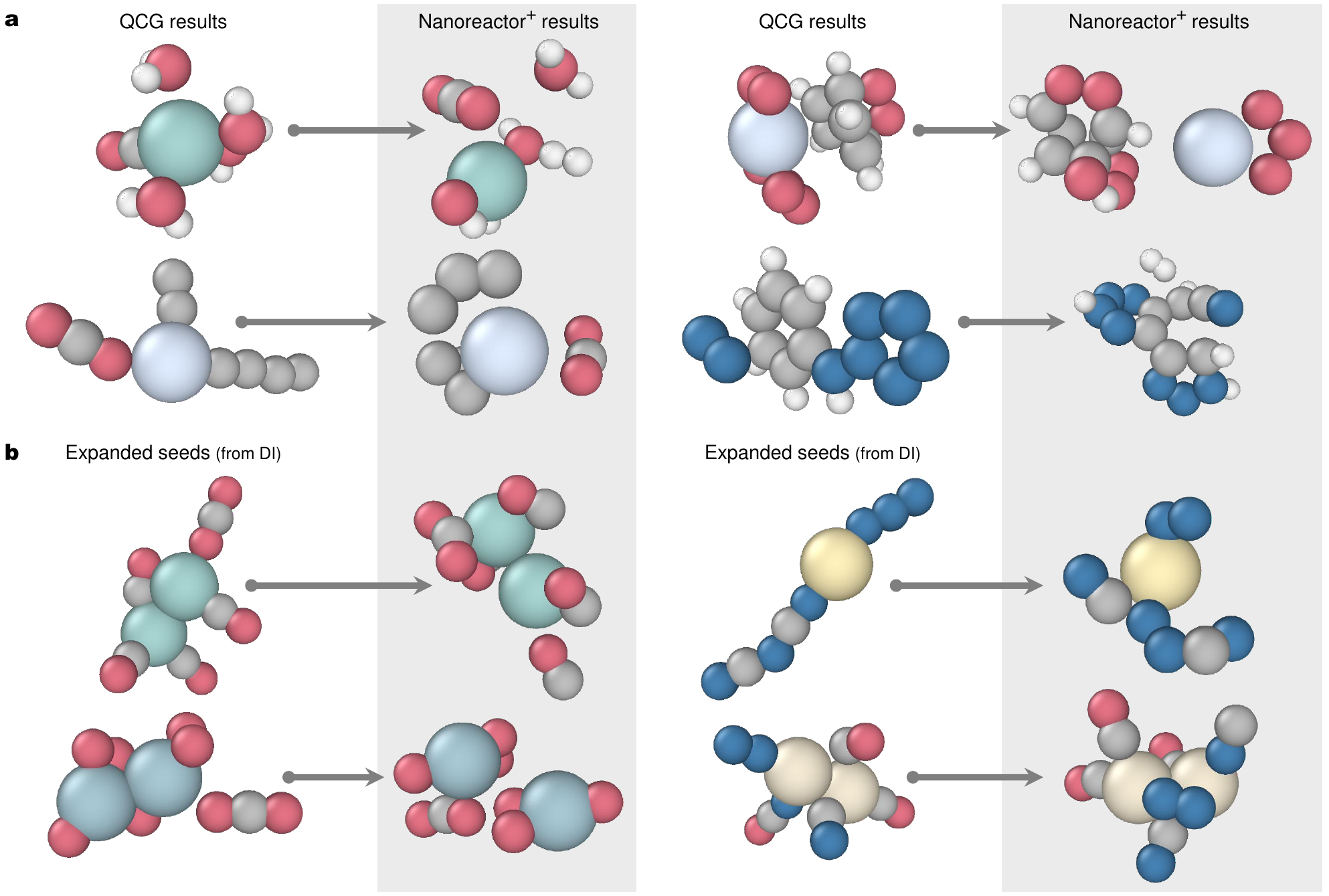}
    \caption{\small
        \textbf{a},~Four representative transformations from input (QCG results) to output (Nanoreactor$^+$ results), indicated by arrows, during the data-generator stage. 
Spheres are color-coded by element; larger spheres represent metals, whereas smaller spheres represent organic elements.
        \textbf{b},~As in \textbf{a}, but during the data-improver stage. The inputs are expanded seeds, and the outputs are again the Nanoreactor$^+$ results.
    }
    \label{sfig:nano}
\end{figure}

\begin{figure}[ht]
    \centering
    \includegraphics[width=1.0\textwidth]{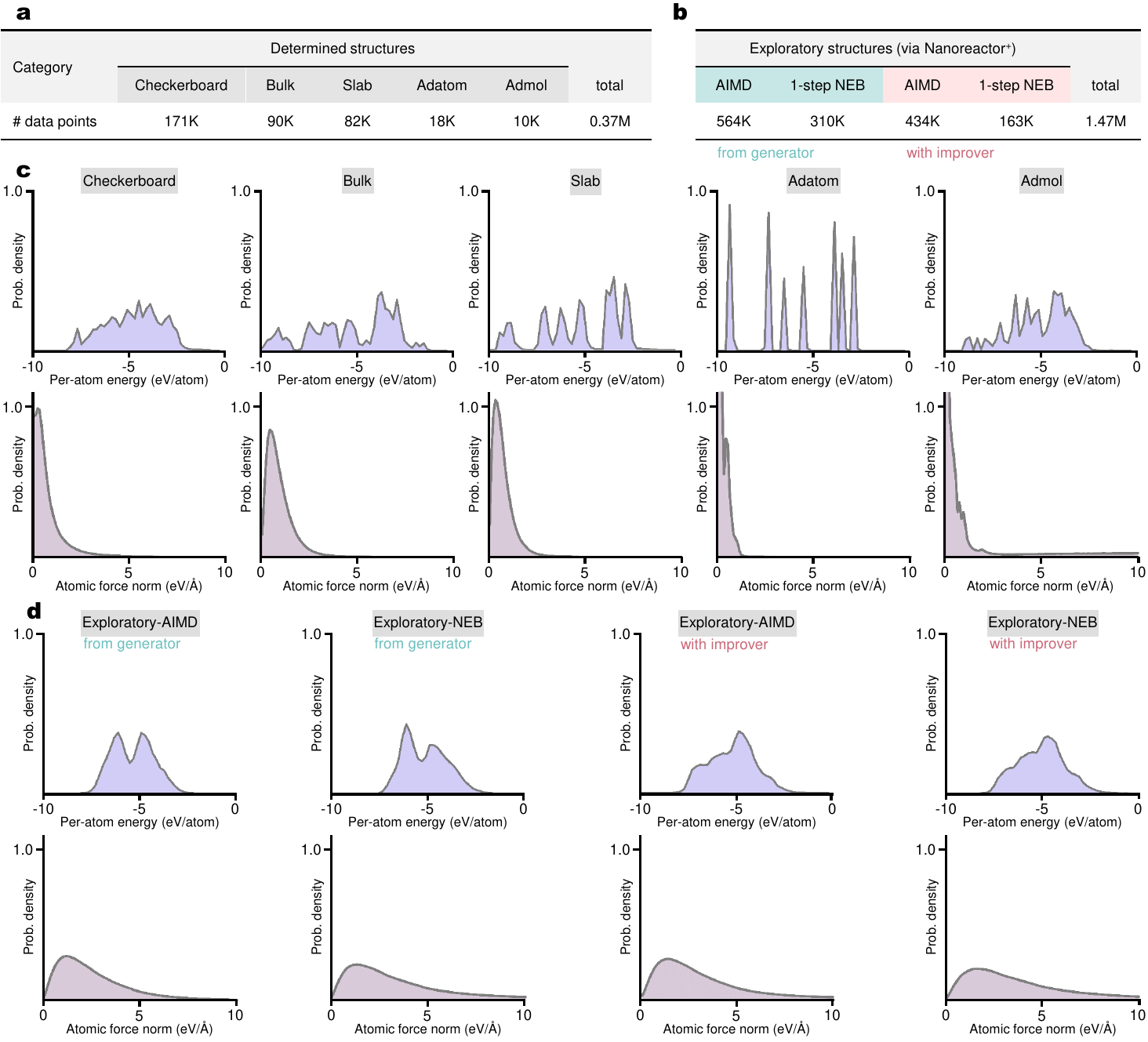}
    \caption{\small
        \textbf{a},~Number of labeled configurations in the Titan25 dataset for each builder (excluding the Nanoreactor$^+$), collectively termed determined structures.
        \textbf{b},~Total number of labeled configurations generated from Nanoreactor$^+$, including those from the data-generator and the data-improver, designated as exploratory structures.
        \textbf{c},~Probability densities of per-atom energy (upper) and atomic force norm (lower) of determined structures, grouped into five pairs. Builder names are annotated at the center of each plot.
        \textbf{d},~As in \textbf{c}, but with exploratory structures generated by the data-generator (left two pairs) and the data-improver (right two pairs). Results labeled by AIMD and NEB are shown separately.
    }
    \label{sfig:titan25}
\end{figure}

\clearpage
\subsection{Model training}\label{supp:mt}
\begin{table}[ht]
\centering
\captionof{table}{\small Hyperparameter details}
\small
\begin{tabular}{ccccc}
\toprule
\multicolumn{1}{c}{Irreps of embedding blocks}          & \multicolumn{4}{c}{128$\times$0e} \\
\multicolumn{1}{c}{Number of interaction blocks}        & \multicolumn{4}{c}{3} \\
\multicolumn{1}{c}{Irreps of interaction blocks}        & \multicolumn{4}{c}{128$\times$0e+64$\times$1e+32$\times$2e+32$\times$3e} \\
                                                        & \multicolumn{4}{c}{128$\times$0e+64$\times$1e+32$\times$2e+32$\times$3e} \\
                                                        & \multicolumn{4}{c}{128$\times$0e} \\
\multicolumn{1}{c}{Cutoff radius (\AA)}                 & \multicolumn{4}{c}{6} \\
\multicolumn{1}{c}{Cutoff function}                     & \multicolumn{4}{c}{XPLOR ~\cite{AGG20}} \\
\multicolumn{1}{c}{Radial basis function}               & \multicolumn{4}{c}{Bessel} \\
\multicolumn{1}{c}{Number of radial basis functions}    & \multicolumn{4}{c}{8} \\
\multicolumn{1}{c}{Energy scaled by}                    & \multicolumn{4}{c}{Root-mean-square of forces} \\
\multicolumn{1}{c}{Energy shifted by}                   & \multicolumn{4}{c}{Per-atom energy mean} \\
\multicolumn{1}{c}{Optimizer}                           & \multicolumn{4}{c}{AdamW} \\
\multicolumn{1}{c}{Batch size}                          & \multicolumn{4}{c}{256} \\
\multicolumn{1}{c}{Learning rate schedule}              & \multicolumn{4}{c}{Cosine} \\
\multicolumn{1}{c}{Warmup epochs}                       & \multicolumn{4}{c}{10\% of training} \\
\multicolumn{1}{c}{Maximum learning rate}               & \multicolumn{4}{c}{0.01} \\
\multicolumn{1}{c}{Gradient clipping}                   & \multicolumn{4}{c}{100} \\
\multicolumn{1}{c}{Weight decay}                        & \multicolumn{4}{c}{0.001} \\
\multicolumn{1}{c}{Energy loss (E)}                     & \multicolumn{4}{c}{Per-atom MAE} \\
\multicolumn{1}{c}{Force loss (F)}                      & \multicolumn{4}{c}{L2MAE ~\cite{KNKC23}} \\
\multicolumn{1}{c}{Stress loss (S)}                     & \multicolumn{4}{c}{MAE} \\
\multicolumn{1}{c}{Loss coefficients}                   & \multicolumn{4}{c}{E:F:S=1:1:1} \\
                            & & & & \\
Datasets                    & ANI-1xnr  & MPTrj & Titan25(G)    & Titan25(G+I) \\ \cmidrule(lr){2-5}
Training steps              & 167K      & 493K  & 486K          & 498K \\
Training epochs             & 300       & 100   & 125           & 80 \\
\bottomrule
\end{tabular}
\label{stab:model_training}
\end{table}

\clearpage
\subsection{Validation performance on Titan25}\label{supp:titan25_valid}

\begin{figure}[ht]
    \centering
    \includegraphics[width=1.0\textwidth]{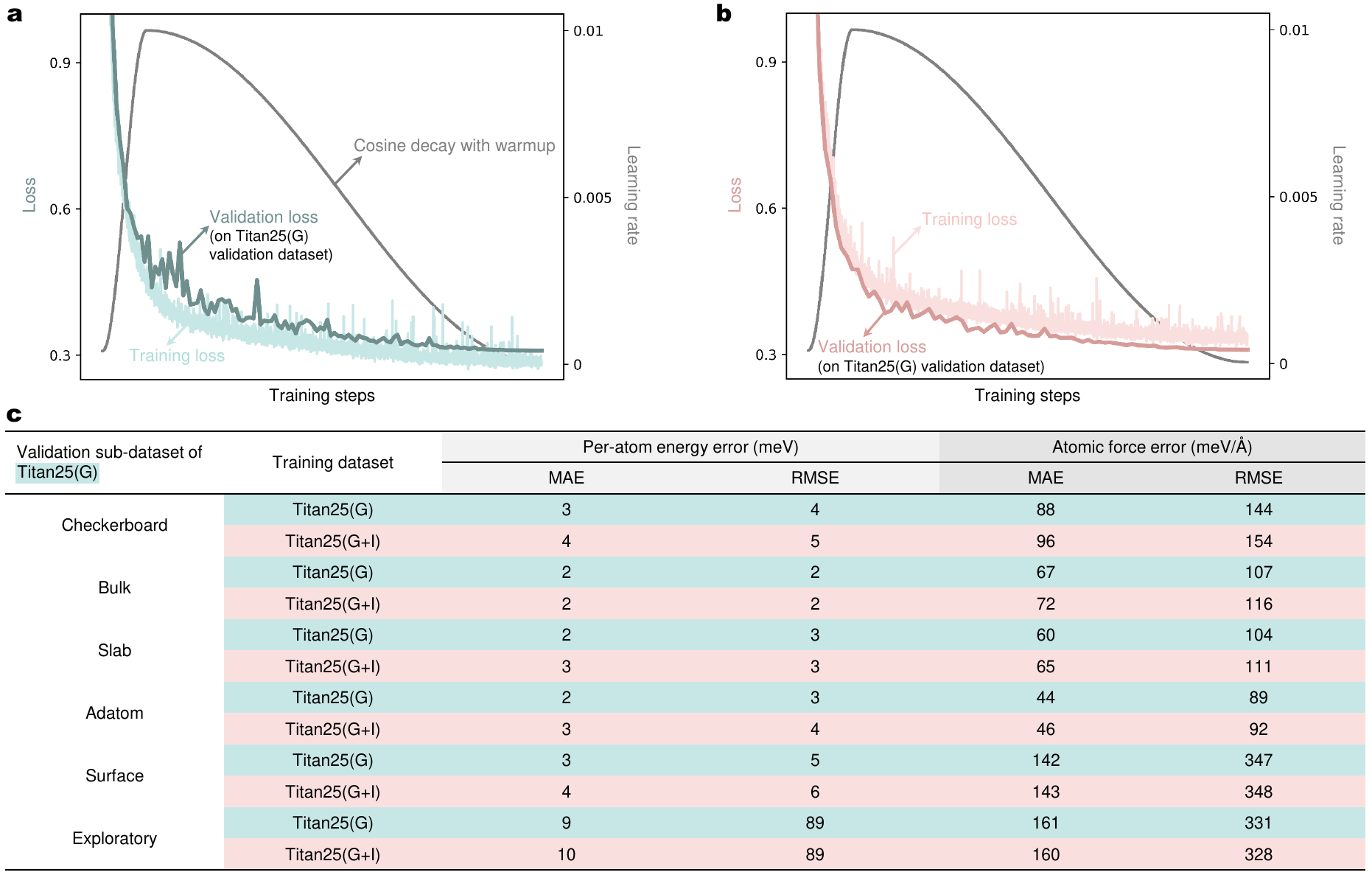}
    \caption{\small
        \textbf{a},\textbf{b},~Validation and training losses on Titan25(G) and Titan25(G+I), respectively.
Learning rates are plotted together over training steps.
        \textbf{c},~Per-atom energy and atomic force errors of validation structures, separated by builder. 
The validation data of Titan25(G) were used in common for both trained models.
    }
    \label{sfig:loss}
\end{figure}

\begin{figure}[ht]
    \centering
    \includegraphics[width=0.8\textwidth]{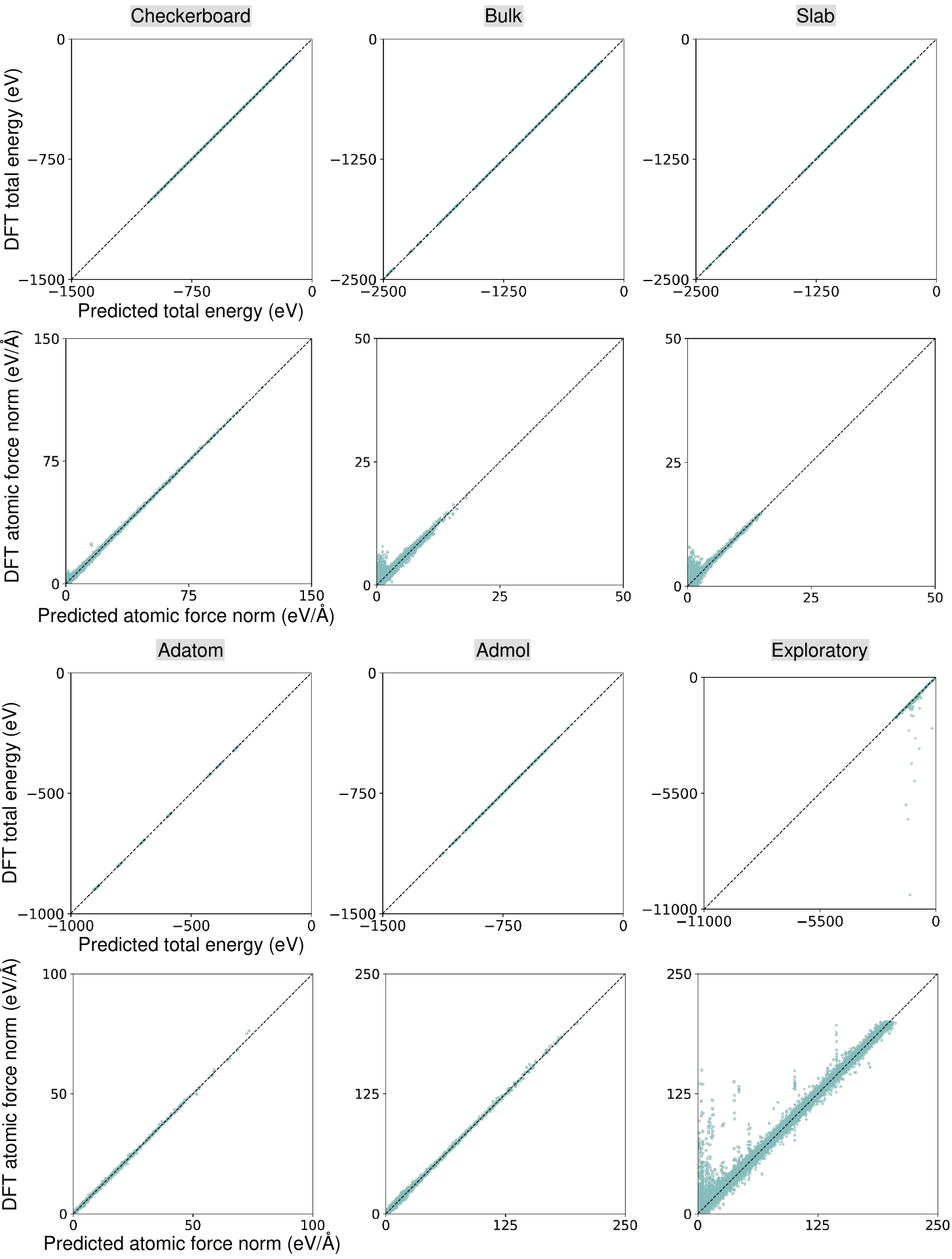}
    \caption{\small
        Parity plots of predictions of the MLIP trained on Titan25(G), against DFT total energies (top) and atomic forces (bottom). 
        They are grouped into six pairs according to their originating builders. 
        The validation data of Titan25(G) were used for the plots.
    }
    \label{sfig:parity_valid}
\end{figure}

\begin{figure}[ht]\ContinuedFloat
    \centering
    \includegraphics[width=0.8\textwidth]{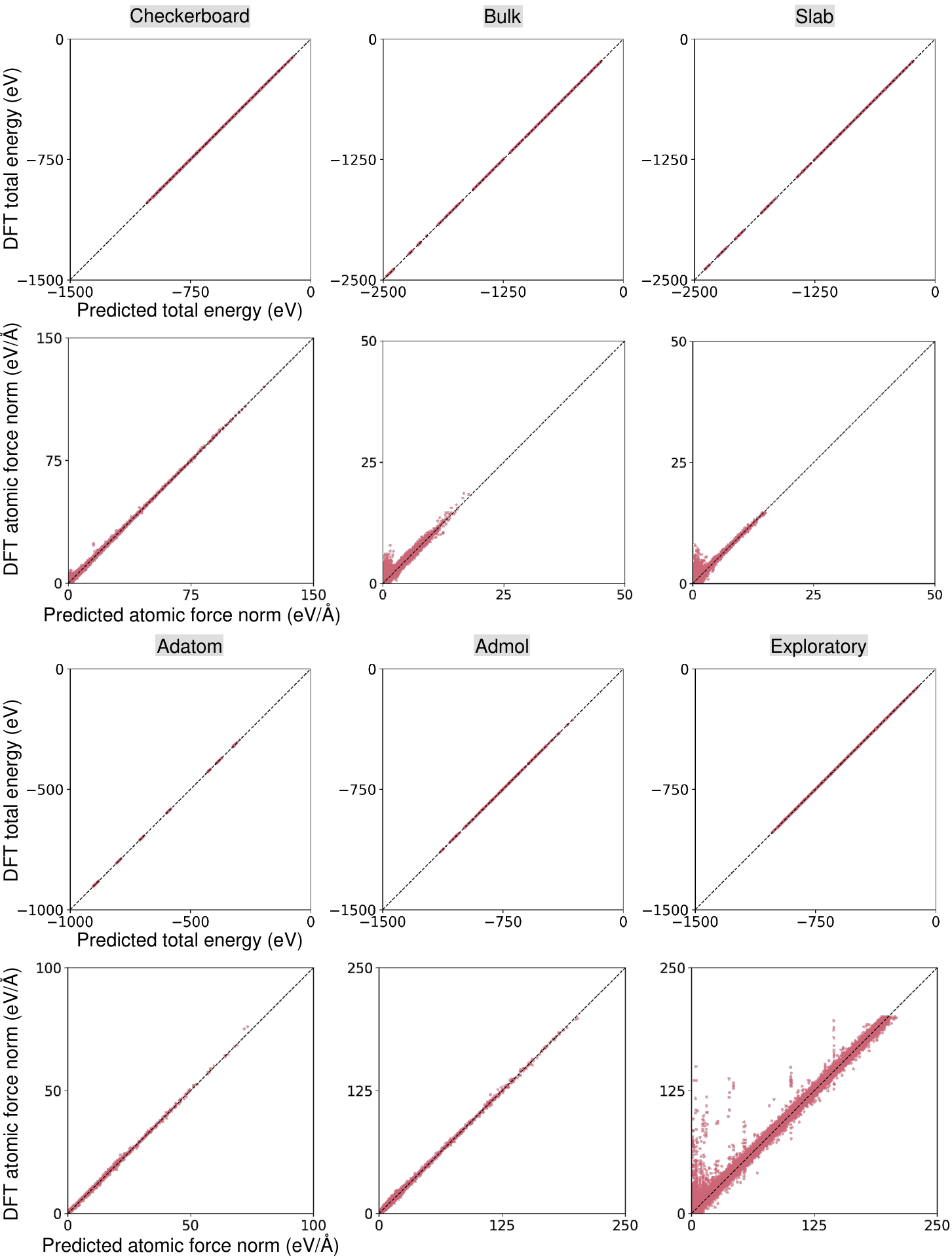}
    \caption{\small (continued)
        Parity plots of predictions of the SNet-T25 model, trained on Titan25(G+I), 
        against DFT total energies (top) and atomic forces (bottom).
    }
\end{figure}

\clearpage
\subsection{GAIA-Bench}\label{supp:gaia_bench}

\begin{figure}[ht]
    \centering
    \includegraphics[width=1.0\textwidth]{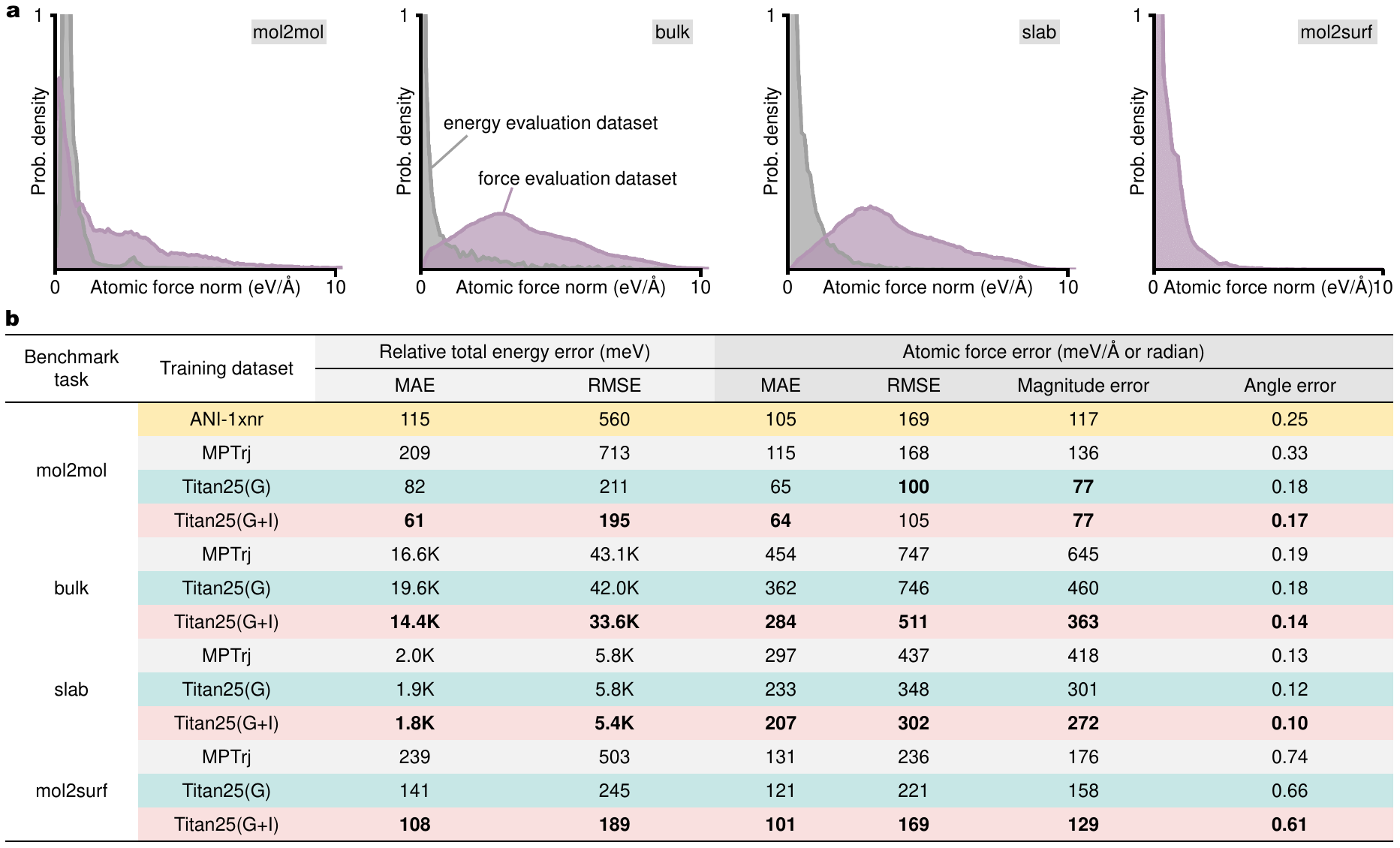}
    \caption{\small
        \textbf{a},~Probability density distributions of the GAIA-Bench tasks with respect to atomic force norm. 
Configurations sampled for force evaluations exhibit a greater weight at larger force norms 
compared to those sampled for energy evaluations, as designed.
        \textbf{b},~MAE and RMSE of relative energies and atomic forces on the four benchmark tasks. 
For forces, the averaged errors of the force magnitude and the angular deviation (in radians) are also reported. 
Boldface indicates the lowest error on each benchmark.
    }
    \label{sfig:bench}
\end{figure}

\clearpage
\begin{figure}[ht]
    \centering
    \includegraphics[width=1.0\textwidth]{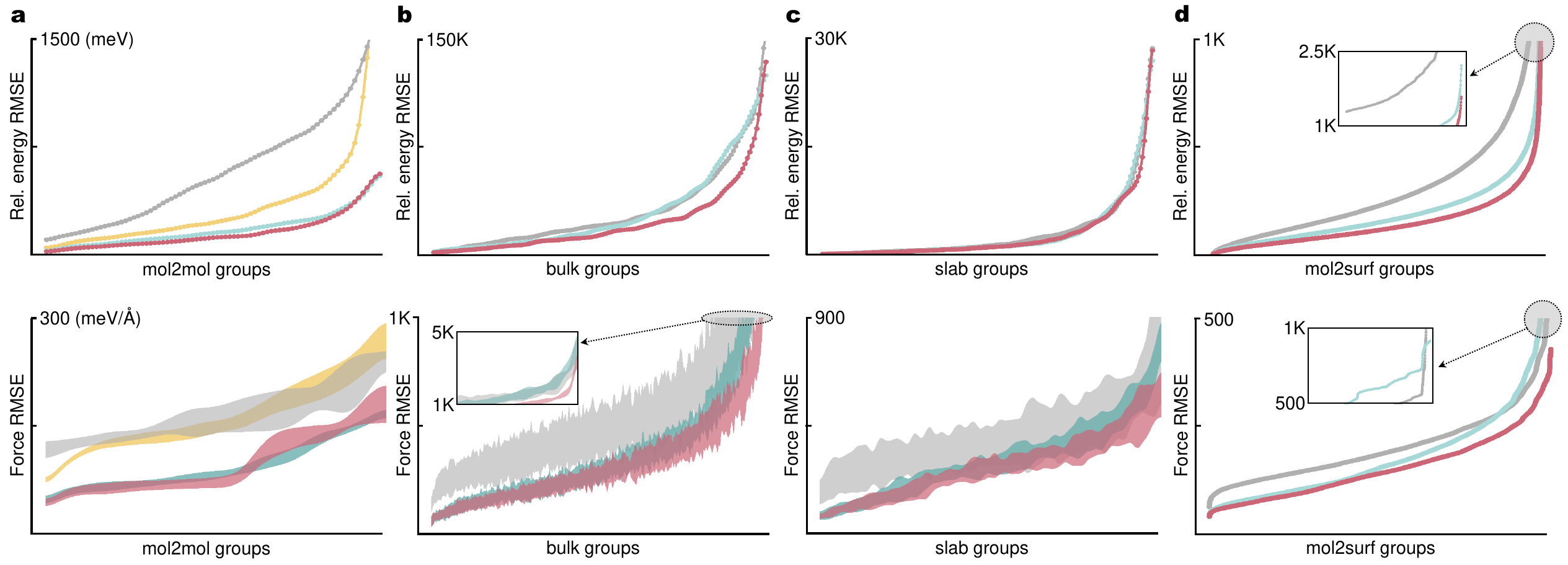}
    \caption{\small
        \textbf{a–d},~Error distributions of relative energies (top) and forces (bottom) for the four GAIA-Bench tasks, 
        as in Fig.~\ref{fig:fig3} but with RMSE.
    }
    \label{sfig:brmse}
\end{figure}

\clearpage
\begin{figure}[ht]
    \centering
    \includegraphics[width=1.0\textwidth]{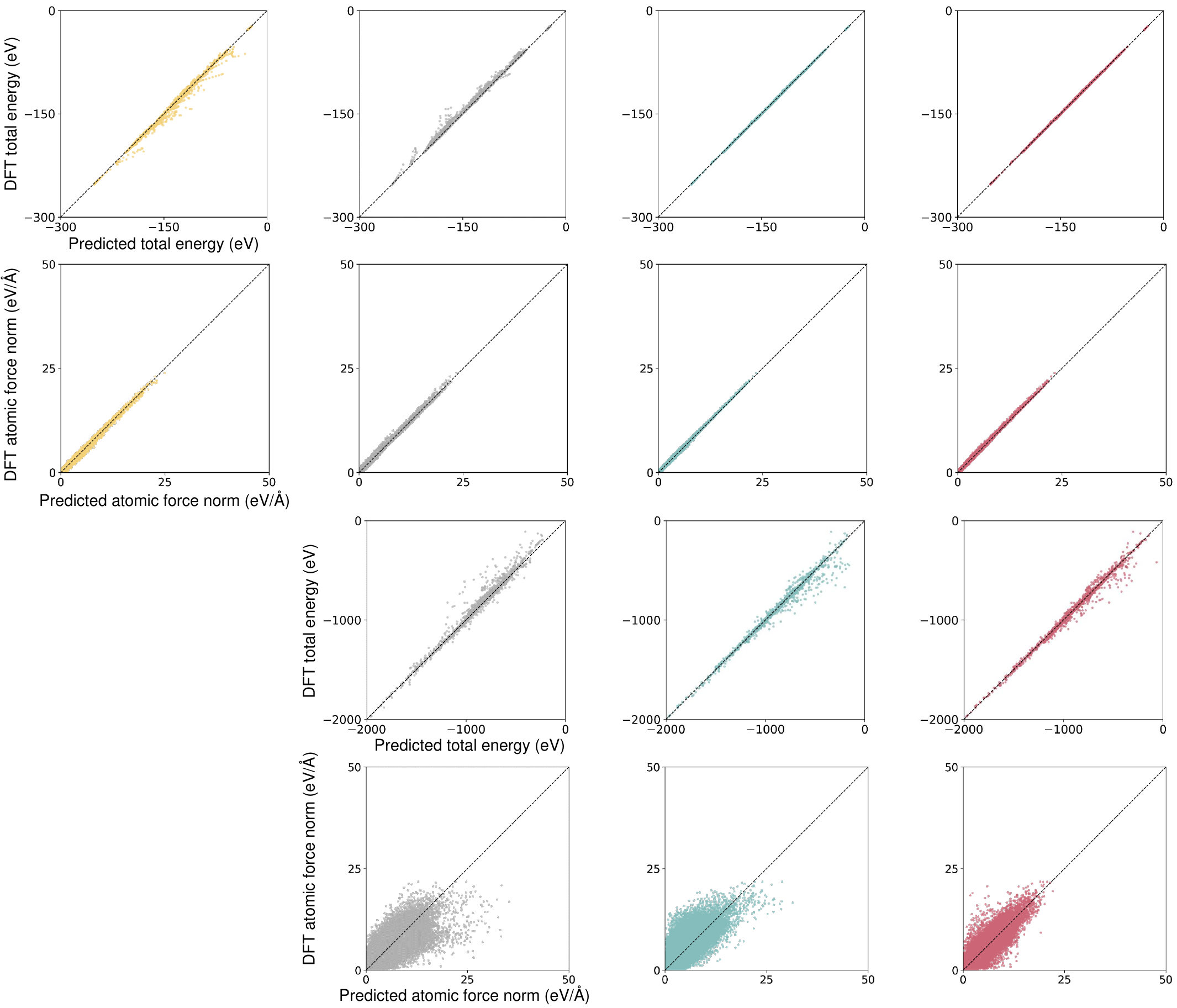}
    \caption{\small
        Parity plots of GAIA-Bench comparing MLIP predictions with reference DFT total energies and atomic force norms. 
        Each column includes the results of the MLIPs trained on ANI-1xnr (yellow), MPTrj (gray), Titan25(G) (blue), and Titan25(G+I) (red), respectively. 
        For ANI-1xnr, results are reported only for mol2mol, 
        as its elemental coverage lacks the metallic species required for the other benchmark tasks.
        Rows correspond to the mol2mol and bulk benchmark results, 
        each shown as a pair of plots for total energy (top) and atomic force norm (bottom).
    }
    \label{sfig:parity_bench}
\end{figure}

\begin{figure}[ht]\ContinuedFloat
    \centering
    \includegraphics[width=1.0\textwidth]{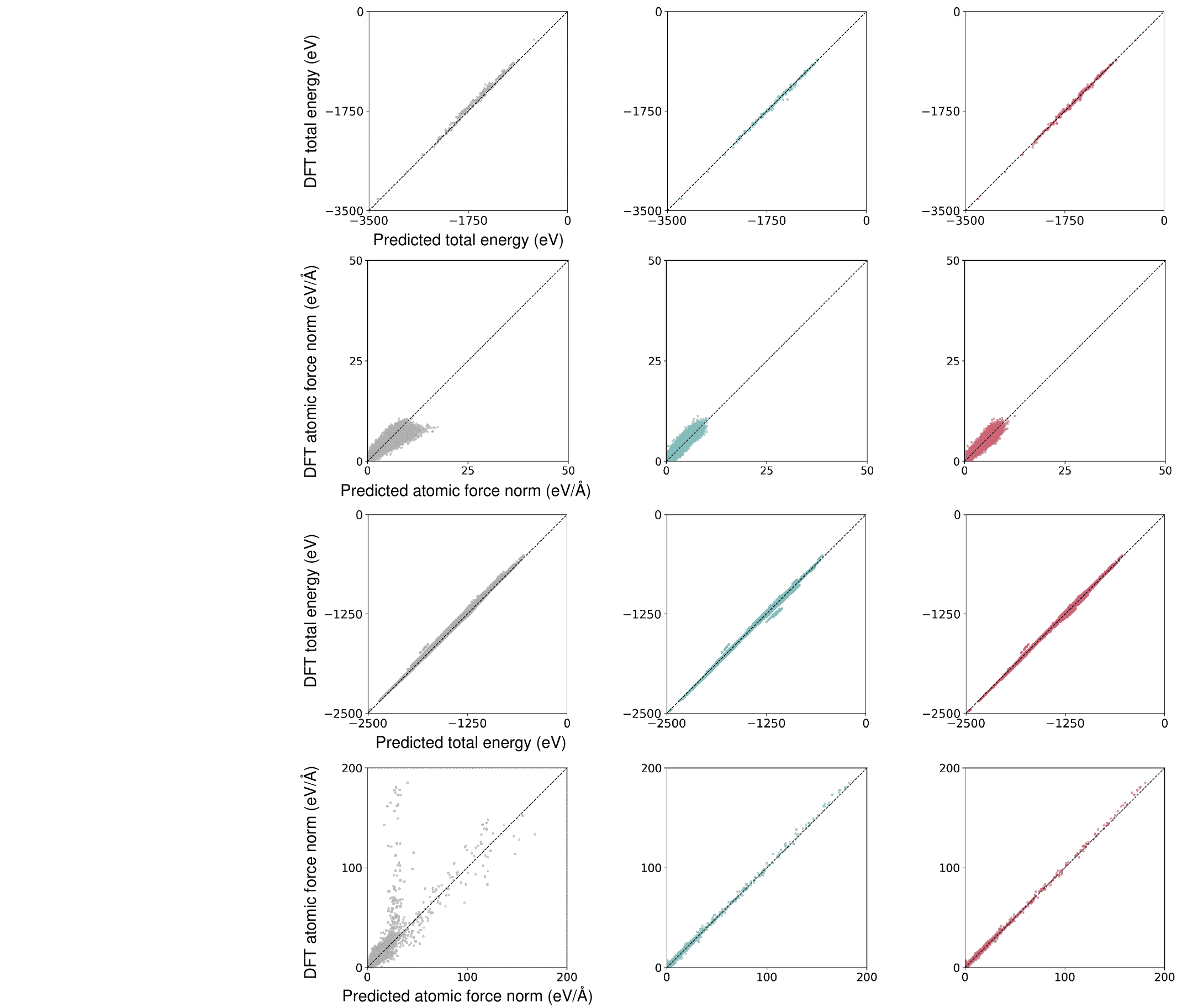}
    \caption{\small (continued)
        Rows correspond to the slab and mol2surf benchmark results, 
        each shown as a pair of plots for total energy (top) and atomic force norm (bottom).
    }
\end{figure}

\clearpage
\begin{table}[ht]
\centering
\captionof{table}{\small
    MAE and RMSE of atomic forces, as in Supplementary Fig.~\ref{sfig:bench}b, 
    but using the energy-evaluated configurations in this test.
    Boldface highlights the lowest error on each benchmark.
}
\small
\begin{tabular}{m{3cm} m{3cm} c c c c}
\toprule
Benchmark task  & Training dataset  & \multicolumn{4}{c}{Atomic force error (meV/\AA\, or radian)} \\
                                    \cmidrule(lr){3-6}
                &                   & MAE & RMSE & Magnitude error & Angle error \\
\midrule
mol2mol         & ANI-1xnr & 198 & 1.6K & 274 & 0.36 \\
                & MPTrj & 210 & 1.3K & 299 & 0.47 \\
                & \cellcolor[HTML]{C7E7E6}Titan25(G) & \cellcolor[HTML]{C7E7E6}84 & \cellcolor[HTML]{C7E7E6}238 & \cellcolor[HTML]{C7E7E6}\textbf{98} & \cellcolor[HTML]{C7E7E6}0.22 \\
                & \cellcolor[HTML]{F9E0DF}Titan25(G+I) & \cellcolor[HTML]{F9E0DF}\textbf{82} & \cellcolor[HTML]{F9E0DF}\textbf{219} & \cellcolor[HTML]{F9E0DF}99 & \cellcolor[HTML]{F9E0DF}\textbf{0.21} \\
\midrule
bulk            & MPTrj & 109 & 410 & 196 & \textbf{1.14} \\

                & \cellcolor[HTML]{C7E7E6}Titan25(G) & \cellcolor[HTML]{C7E7E6}130 & \cellcolor[HTML]{C7E7E6}503 & \cellcolor[HTML]{C7E7E6}223 & \cellcolor[HTML]{C7E7E6}1.17 \\
                & \cellcolor[HTML]{F9E0DF}Titan25(G+I) & \cellcolor[HTML]{F9E0DF}\textbf{92} & \cellcolor[HTML]{F9E0DF}\textbf{304} & \cellcolor[HTML]{F9E0DF}\textbf{160} & \cellcolor[HTML]{F9E0DF}\textbf{1.14} \\
\midrule
slab            & MPTrj & 131 & 219 & 176 & 0.82 \\
                & \cellcolor[HTML]{C7E7E6}Titan25(G) & \cellcolor[HTML]{C7E7E6}145 & \cellcolor[HTML]{C7E7E6}263 & \cellcolor[HTML]{C7E7E6}189 & \cellcolor[HTML]{C7E7E6}0.80 \\
                & \cellcolor[HTML]{F9E0DF}Titan25(G+I) & \cellcolor[HTML]{F9E0DF}\textbf{119} & \cellcolor[HTML]{F9E0DF}\textbf{195} & \cellcolor[HTML]{F9E0DF}\textbf{152} & \cellcolor[HTML]{F9E0DF}\textbf{0.76} \\
\midrule
mol2surf        & MPTrj & 131 & 236 & 176 & 0.74 \\
                & \cellcolor[HTML]{C7E7E6}Titan25(G) & \cellcolor[HTML]{C7E7E6}121 & \cellcolor[HTML]{C7E7E6}221 & \cellcolor[HTML]{C7E7E6}158 & \cellcolor[HTML]{C7E7E6}0.66 \\
                & \cellcolor[HTML]{F9E0DF}Titan25(G+I)  & \cellcolor[HTML]{F9E0DF}\textbf{101} & \cellcolor[HTML]{F9E0DF}\textbf{169} & \cellcolor[HTML]{F9E0DF}\textbf{129} & \cellcolor[HTML]{F9E0DF}\textbf{0.61} \\
\bottomrule
\end{tabular}
\label{stab:bench_e2f}
\end{table}

\clearpage
\subsection{Test performance on public datasets}\label{supp:public_test}

\begin{table}[ht]
\centering
\captionof{table}{\small
    Energy and atomic force error comparison of four MLIP models on public test datasets: 
    ANI-1x~\cite{SZNL20}, GDB-13~\cite{BR09}, ANI-1xnr~\cite{ZMJK24s}, and MPTrj~\cite{DZJR23s}. 
    All datasets were re-labeled with the PBE-D3 functional for consistency in the comparison. 
    For the test datasets, ANI-1x and GDB-13 were evaluated by randomly sampling 20,000 data points. 
    In contrast, ANI-1xnr and MPTrj were partitioned into training (80\%), validation (10\%), and test (10\%), 
    with the model performance assessed on the held-out 10\% validation sets to allow fair comparison with models trained on the corresponding training datasets.
    The validation set of MPTrj was further restricted to configurations containing only elements present in Titan25. 
    Entries corresponding to the lowest errors in each test dataset are highlighted in bold.
}
\small
\begin{tabular}{m{2cm} m{3cm} c c c c}
\toprule
Test dataset & Training dataset  & \multicolumn{2}{c}{Per-atom energy error (meV)} & \multicolumn{2}{c}{Atomic force error (meV/\AA)} \\
                                    \cmidrule(lr){3-4}\cmidrule(lr){5-6}
             &                   & MAE & RMSE & MAE & RMSE \\
\midrule
ANI-1x       & ANI-1xnr & 50 & 56 & 230 & 363 \\
             & MPTrj & 35 & 44 & 290 & 363 \\
             & \cellcolor[HTML]{C7E7E6}Titan25(G) & \cellcolor[HTML]{C7E7E6}\textbf{27} & \cellcolor[HTML]{C7E7E6}\textbf{31} & \cellcolor[HTML]{C7E7E6}143 & \cellcolor[HTML]{C7E7E6}226 \\
             & \cellcolor[HTML]{F9E0DF}Titan25(G+I) & \cellcolor[HTML]{F9E0DF}\textbf{27} & \cellcolor[HTML]{F9E0DF}32 & \cellcolor[HTML]{F9E0DF}\textbf{137} & \cellcolor[HTML]{F9E0DF}\textbf{217} \\
\midrule
GDB-13       & ANI-1xnr & 36 & 41 & 186 & 336 \\
             & MPTrj & \textbf{17} & 24 & 190 & 319 \\
             & \cellcolor[HTML]{C7E7E6}Titan25(G) & \cellcolor[HTML]{C7E7E6}23 & \cellcolor[HTML]{C7E7E6}25 & \cellcolor[HTML]{C7E7E6}108 & \cellcolor[HTML]{C7E7E6}182 \\
             & \cellcolor[HTML]{F9E0DF}Titan25(G+I) & \cellcolor[HTML]{F9E0DF}20 & \cellcolor[HTML]{F9E0DF}\textbf{23} & \cellcolor[HTML]{F9E0DF}\textbf{101} & \cellcolor[HTML]{F9E0DF}\textbf{167} \\
\midrule
ANI-1xnr     & ANI-1xnr & 4 & 6 & 139 & 251 \\
             & MPTrj & 54 & 68 & 489 & 817 \\
             & \cellcolor[HTML]{C7E7E6}Titan25(G) & \cellcolor[HTML]{C7E7E6}13 & \cellcolor[HTML]{C7E7E6}15 & \cellcolor[HTML]{C7E7E6}243 & \cellcolor[HTML]{C7E7E6}373 \\
             & \cellcolor[HTML]{F9E0DF}Titan25(G+I) & \cellcolor[HTML]{F9E0DF}12 & \cellcolor[HTML]{F9E0DF}14 & \cellcolor[HTML]{F9E0DF}232 & \cellcolor[HTML]{F9E0DF}354 \\
\midrule
MPTrj        & MPTrj & 38 & 77 & 96 & 248 \\
             & \cellcolor[HTML]{C7E7E6}Titan25(G) & \cellcolor[HTML]{C7E7E6}107 & \cellcolor[HTML]{C7E7E6}174 & \cellcolor[HTML]{C7E7E6}132 & \cellcolor[HTML]{C7E7E6}272 \\
             & \cellcolor[HTML]{F9E0DF}Titan25(G+I) & \cellcolor[HTML]{F9E0DF}83 & \cellcolor[HTML]{F9E0DF}126 & \cellcolor[HTML]{F9E0DF}112 & \cellcolor[HTML]{F9E0DF}234 \\
\bottomrule
\end{tabular}
\label{stab:public_test}
\end{table}

\clearpage
\subsection{MLIP-MD evaluations}\label{supp:mlip_md}

\begin{figure}[ht]
    \centering
    \includegraphics[width=0.85\textwidth]{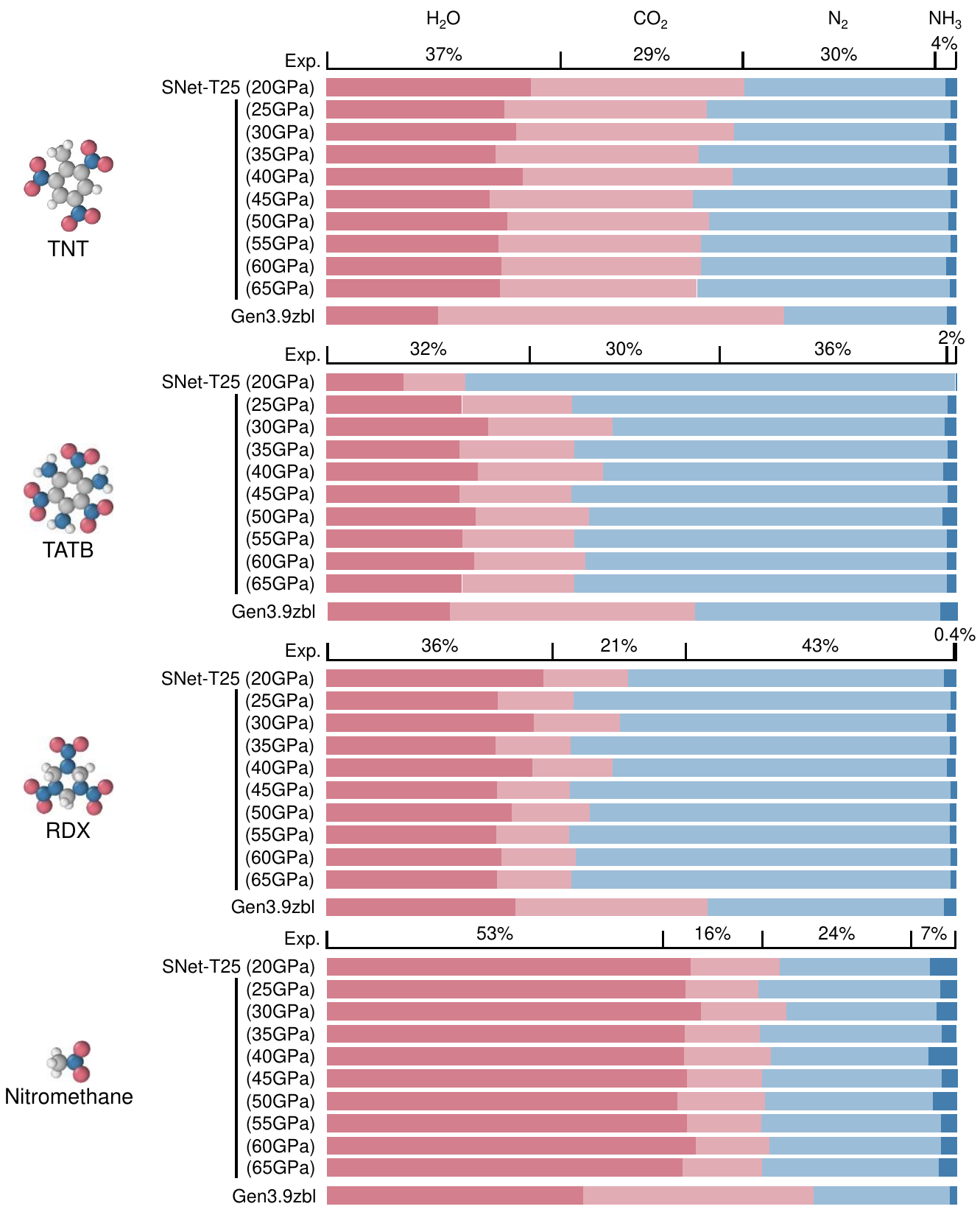}
    \caption{\small 
        Normalized product ratios of the four energetic molecules simulated with the SNet-T25 model. 
        Gray, white, blue, and red spheres represent C, H, N, and O atoms, respectively. 
        Results for Gen3.9zbl are taken from ref.~\cite{HYSI23s}.
    }
    \label{sfig:deto}
\end{figure}

\clearpage
\begin{figure}[ht]
    \centering
    \includegraphics[width=1.0\textwidth]{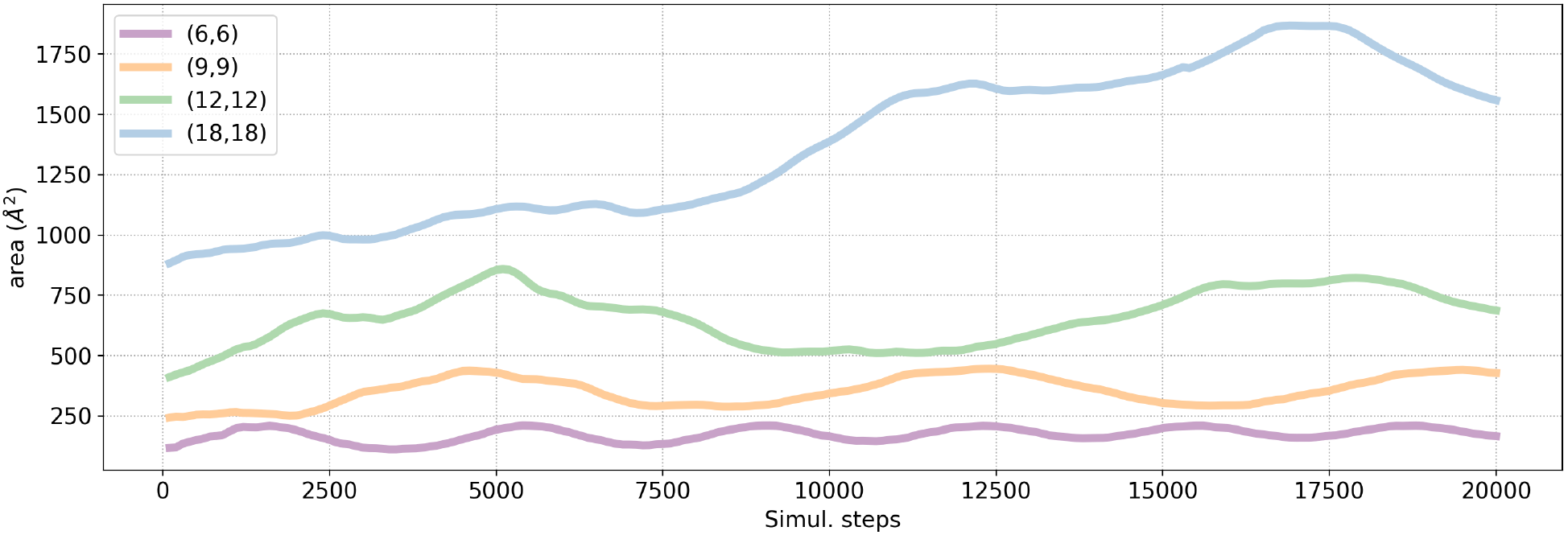}
    \caption{\small
        Projected area of CNTs with chirality indices (6,6), (9,9), (12,12), and (18,18) plotted against simulation steps. 
        The fluctuations in cross-sectional area, observed in each curve, 
        arise from structural transformations of the CNTs induced by thermal effects.
    }
    \label{sfig:cnt}
\end{figure}

\clearpage
\begin{figure}[ht]
    \centering
    \includegraphics[width=1.0\textwidth]{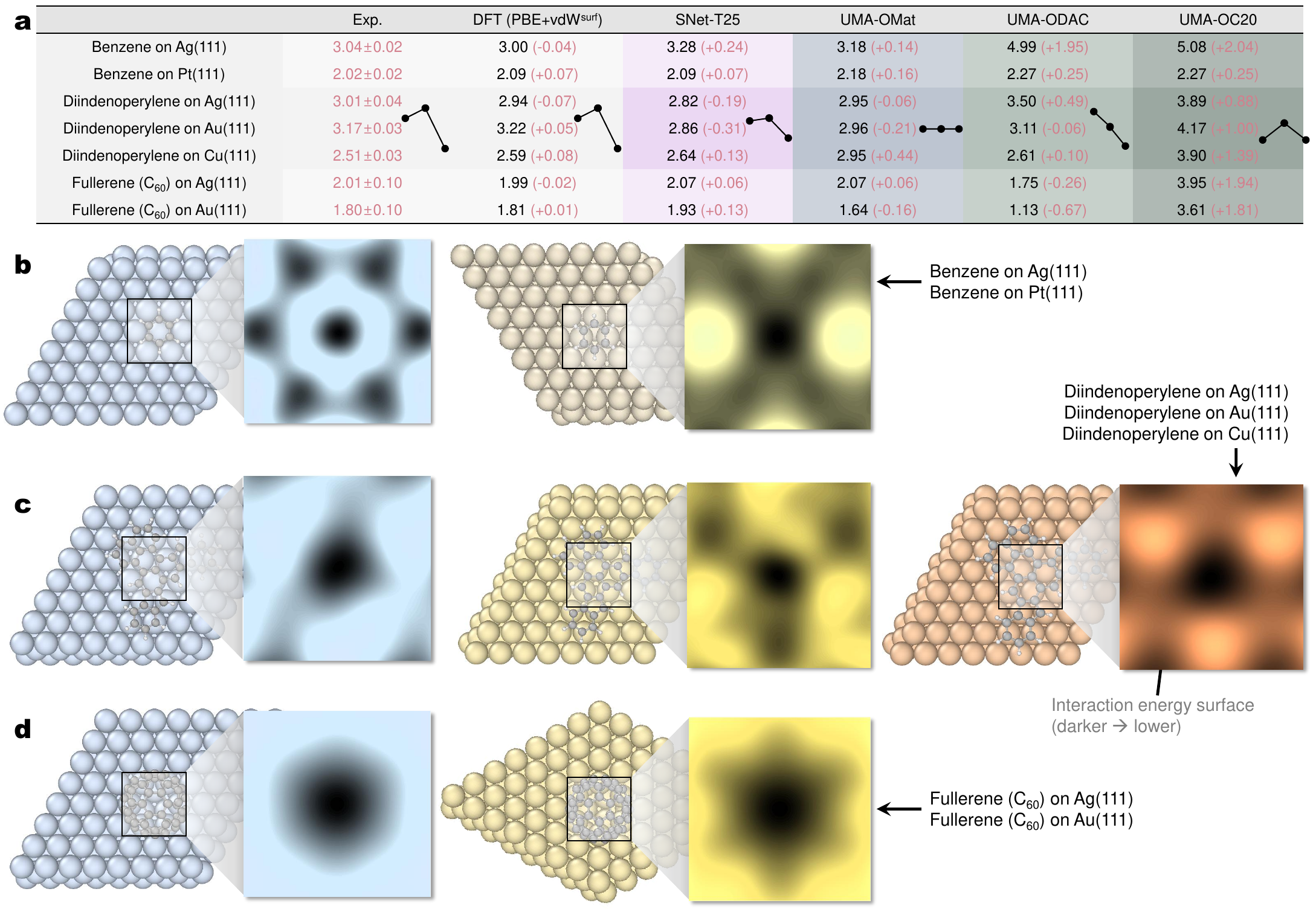}
    \caption{\small
        \textbf{a},~Vertical distances of $\pi$-conjugated molecules in their relaxed structures on various metallic surfaces, obtained with different MLIPs. 
The $\pm$ values indicate error bars for the experiments; 
the values in parentheses denote deviations from experiment. 
In the diindenoperylene rows, the trend of distance variations across the materials is also described next to the values.
SNet-T25 and UMA-OMat24 reproduce the experimental trend well. All quantities are reported in \AA. 
        \textbf{b},~Benzene on Ag(111) and Pt(111).
        \textbf{c},~Diindenoperylene on Ag(111), Au(111), and Cu(111).
        \textbf{d},~Fullerene on Ag(111) and Au(111). Note that there is a central vacancy on the metallic surface.
    }
    \label{sfig:pi-conj-relaxation}
\end{figure}

\clearpage
\begin{figure}[ht]
    \centering
    \includegraphics[width=1.0\textwidth]{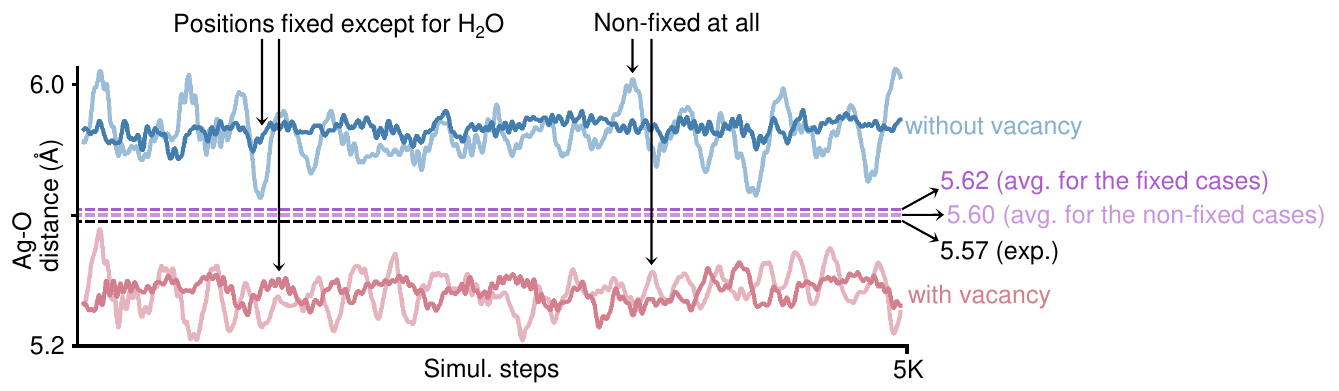}
    \caption{\small
        Evolution of the Ag--O distance during MD simulations of a fullerene on the Ag(111) surface. 
        The red curves correspond to the fullerene encapsulating H$_2$O on Ag(111) with a central vacancy, 
        whereas the blue curves represent the defect-free Ag(111) surface. 
        Light-colored curves indicate unconstrained MD simulations; 
        solid-colored curves denote simulations in which all atoms were fixed except for H$_2$O. 
        The average Ag--O distances are shown as dashed lines in pale purple and purple, respectively.
    }
    \label{sfig:pi-conj-md}
\end{figure}

\clearpage




\end{document}